\documentclass[useAMS,usenatbib]{mnras}
\pdfoutput=1
\usepackage{graphicx,psfig,cite,epsf}  
\usepackage{times}
\usepackage{subfigure}
\usepackage{multirow}
\usepackage{amsmath}	
\usepackage{amssymb}	
\usepackage{gensymb}
\usepackage{colortbl}
\usepackage[utf8]{inputenc}
\usepackage{newtxtext,newtxmath}
\usepackage{color}
\usepackage{graphicx}
\usepackage{natbib}
\usepackage{cite}
\usepackage{geometry}
\usepackage{ae,aecompl}
\usepackage{longtable}
\usepackage{soul}
\usepackage{lscape}

\usepackage{array,etoolbox}
\preto\tabular{\setcounter{magicrownumbers}{0}}
\newcounter{magicrownumbers}
\def\rownumber{}

\date{Accepted XXX. Received YYY; in original form ZZZ}

\pubyear{2021}

\def\swift{{\it Swift}/XRT}
\def\xmm{{\it XMM-Newton}}
\def\chandra{{\it Chandra}}
\def\nus{{\it NuSTAR}}

\title[Ultraluminous X-ray sources in the galaxy NGC 925]{Investigating the nature of the ultraluminous X-ray sources in the galaxy NGC 925}

\author[Chiara Salvaggio et al.]{
Chiara Salvaggio,$^{1,2}$\thanks{E-mail: chiara.salvaggio@inaf.it}
A. Wolter,$^{2}$\thanks{E-mail: anna.wolter@inaf.it}
F. Pintore,$^{3}$
C. Pinto,$^{3}$
E. Ambrosi,$^{3}$
G. L. Israel,$^{4}$
\newauthor
A. Marino,$^{5,3}$
R. Salvaterra,$^{6}$
L. Zampieri,$^{7}$
A. Belfiore$^{6}$
\\
$^{1}$Dipartimento di Fisica, Universit\`a degli Studi di Milano - Bicocca, Piazza della Scienza 3, 20126 Milano, Italy\\
$^{2}$INAF - Osservatorio Astronomico di Brera, via Brera 28, 20121 Milano, Italy\\
$^{3}$INAF - IASF Palermo, Via U. La Malfa 153, I-90146 Palermo, Italy\\
$^{4}$INAF - Osservatorio astronomico di Roma, Via Frascati 44, I-00040, Monteporzio Catone, Italy\\
$^{5}$Universitá degli Studi di Palermo, Dipartimento di Fisica e Chimica, via Archirafi 36, I-90123 Palermo, Italy\\
$^{6}$INAF - Istituto di Astrofisica Spaziale e Fisica Cosmica di Milano, via A. Corti 12, 20133 Milano, Italy\\
$^{7}$INAF - Osservatorio Astronomico di Padova, Vicolo dell’Osservatorio 5, I-35122 Padova, Italy\\
}

\date{April 2021}

\begin{document}
\label{firstpage}
\pagerange{\pageref{firstpage}--\pageref{lastpage}}
\maketitle

\begin{abstract}
Variability is a powerful tool to investigate properties of X-ray binaries (XRB), in particular for Ultraluminous X-ray sources (ULXs)  that are mainly detected in the X-ray band. For most ULXs the nature of the accretor is unknown, although a few ULXs have been confirmed to be accreting at super-Eddington rates onto a neutron star (NS). 
Monitoring these sources is particularly useful both to detect transients and to derive periodicities, linked to orbital and super-orbital modulations. 
Here we present the results of our monitoring campaign of the galaxy NGC 925, performed with the \textit{Neil Gehrels Swift Observatory}. We also include archival and literature data obtained with \chandra, \textit{XMM-Newton} and \nus. We have studied spectra, light-curves and variability properties on days to months time-scales. All the three ULXs detected in this galaxy show flux variability. ULX-1 is one of the most luminous ULXs known, since only 10\% of the ULXs exceed a luminosity of $\sim$5$\times$10$^{40}$ erg s$^{-1}$, but despite its high flux variability we found only weak spectral variability. We classify it as in a hard ultraluminous regime of super-Eddington accretion. ULX-2 and ULX-3 are less luminous but also variable in flux and possibly also in spectral shape. We classify them as in between the hard and the soft ultraluminous regimes. ULX-3 is a transient source: by applying a Lomb-Scargle algorithm we derive a periodicity of $\sim$ 126 d, which could be associated with an orbital or super-orbital origin.
\end{abstract}


\section{Introduction}
\label{sec:introduction}
Ultraluminous X-ray sources (ULXs) are extragalactic, point-like and non-nuclear objects, with luminosities in excess of 10$^{39}$ erg s$^{-1}$, interpreted as accreting compact objects in binary systems (see e.g. \citealt{Kaaret2017} for a review). While at first they were preferentially considered the ideal place to host intermediate mass black holes (IMBH, e.g. \citealt{ColbertMushotzky1999}, \citealt{Ebisuzaki2001}, \citealt{Farrell2009}) accreting at sub-Eddington rates, they are now widely thought to be powered by super-Eddington accretion onto stellar mass compact objects like a neutron star (NS) or a black hole (BH). The ULXs are typically observed in the so called \textit{ultraluminous state} (e.g. \citealt{Roberts2007, Gladstone2009b}), characterised by a spectral curvature at 2$-$5 keV usually coupled to a soft excess peaking in the range 0.1--0.5 keV  \citep[e.g.][]{Gladstone2009b}, and at times by temporal variability on short time-scales \citep[e.g.][]{Heil2009,Sutton2013,pintore2014,middleton2015a}. Confirmation that a fraction of the ULX population contains a stellar-mass compact object comes from the detection of coherent pulsations in six ULXs \citep{Bachetti2014,Israel2017,israel2017b,Fuerst2016,Carpano2018,Rodriguez2019,Sathyaprakash2019}, which can be explained only by the presence of a NS in these systems, and from the detection of a transient cyclotron line in the NS candidate M51 ULX-8 \citep{brightman2018}. Considering that only $\sim$20 sources have enough statistics to detect pulsations (e.g. \citealt{Earnshaw2019}; \citealt{Song2020}), the NSs may be a significant fraction among the bright ULX population (\citealt{Pintore2017,Koliopanos2017,Walton2018a}). 

The short-term (seconds to hours) temporal properties of ULXs are quite different from those observed in Galactic X-ray binary systems and are not related to specific spectral states \citep[e.g.][]{Heil2009}: even if the amplitude of short-term variability is larger in soft ultraluminous sources than in hard ones \citep{Sutton2013}, the variability appears to be poorly predictable, i.e. it is not found in all sources with a soft spectrum and, when detected in a source, it is not necessarily found in all the observations of that ULX. Such variability was explained with the existence of optically thick and non-uniform winds, radiatively ejected by the accretion disc, which stochastically intersect our line of sight (e.g. \citealt{Middleton2011,middleton2015a}). Observational evidences of these winds come from the detection of blue-shifted absorption lines in the high quality grating spectra of some ULXs (e.g. \citealt{pinto2016,kosec2018b}) and by the nebulae observed  in the radio (e.g. \citealt{cseh2012}), 
optical (e.g. \citealt{Pakull2010}) or X-ray band (\citealt{Belfiore2020}). Long-term flux variability (days to months) is also observed in most of the monitored ULXs. Such variability, in some cases, can be as high as several orders of magnitudes, implying that these sources are transients (detected at least once in the ultraluminous state, L$_{x} > $ 10$^{39}$ erg s$^{-1}$, and either once at a significantly smaller luminosity or undetected below the instrumental sensitivity; e.g. \citealt{Pintore2018a,Soria2012}). Often the pulsating ULXs (PULXs) show large flux variations, implying they can be considered as transient ULXs. A bi-modal flux distribution is sometimes observed in the long-term light-curves of PULXs (e.g. \citealt{Walton2015a,motch2014}). A possible explanation for this long-term behaviour may be the propeller effect: when the magnetospheric radius of the NS becomes larger than the corotation radius of the accreting matter in the disc, a centrifugal barrier prevents the accretion onto the NS, with a corresponding decrease in the X-ray flux (\citealt{Illarionov1975,Tsygankov2016a,Grebenev2017}). The propeller effect can act only if the accretor is a NS with a strong magnetic field, so a bi-modal flux distribution can be used to identify candidate PULXs, even when pulsations are not detected (e.g. \citealt{Earnshaw2018,Song2020}). Furthermore, long-term monitored ULXs revealed possible (super-)orbital variability \citep[e.g.][]{Foster2010,An2016,Fuerst2018}, the origin of which is still a matter of debate. 
Super-orbital periods are also observed in the light-curves of the PULXs, with periods of tens to hundreds days \citep[e.g.][]{Walton2016,Brightman2019,Brightman2020}. 
The detection of such long-term periodicities were only possible thanks to the flexibility and performance of the \textit{Neil Gehrels Swift Observatory} (hereafter \textit{Swift}; \citealt{Gehrels2004}).

Some ULXs present also spectral variability (see, for example, \citealt{Bachetti2013,pintore2014}). \citet{Pintore2017} suggested, based on hardness ratio studies, that the known PULXs generally have harder spectra amongst the ULXs population. We use here the spectral classification in three spectral regimes proposed by 
\citet{Sutton2013}, based on the spectral parameters of a multi-colour disc plus a power-law model: \textit{broadened disc, hard ultraluminous} and \textit{soft ultraluminous}. Sources with an inner disc temperature $<$ 0.5 keV are classified in the \textit{hard ultraluminous regime}, if the power-law index is smaller than 2, and in the \textit{soft ultraluminous regime}, if the power-law index is larger than 2. 
When the inner disc temperature assumes values larger than 0.5 keV, if the ratio between the power-law flux and the disc flux in the (0.3--1) keV band is larger than 5, we have again an \textit{ultraluminous regime}, \textit{hard} or \textit{soft} depending on the power-law index. Otherwise, when the flux ratio is smaller than 5, the spectrum is characterised by a dominant disc component, and the regime is labelled as \textit{broadened disc}. The \textit{broadened disc} is characterised by a broad disc-like spectral shape and is very common in the lower-luminosity sources (below $\sim$ 3$\times$10$^{39}$ erg s$^{-1}$) but it is sometimes also observed at higher luminosities. The two ultraluminous regimes are instead characterised by two spectral components, with a peak in the higher (hard ultraluminous) or lower (soft ultraluminous) energy component of the 0.3-10 keV spectrum. 

Such regimes are thought to depend both on the mass accretion rate and on the observer's viewing angle: if the system is seen face-on, the hard inner emission is directly detected (hard-ultraluminous regime), while for more inclined systems the emission from the innermost regions is seen through the wind, appearing softer (soft-ultraluminous regime).  
Both ultraluminous regimes may coexist in a single system. The switching between the two regimes is determined by changes in the accretion rate driving changes in the outflowing wind opening angle (e.g. \citealt{Sutton2013,middleton2015a}).

Most of the X-ray flux variability is observed in the hard band, above 1 keV,  
and could be associated to temporary obscuration of the hot inner disc regions by the funnel created by the puffed disc and the wind (e.g. \citealt{middleton2015a,pinto2017}). However, the rather stable hard flux above 10 keV as observed in some archetypal ULXs like NGC 1313 X-1 and Holmberg IX X-1 suggests that the scenario might be more complicated \citep{Walton2017,Walton2020, Gurpide2021a}. This is confirmed by the findings of different variability patterns such as flaring activity in some hard ULXs (e.g. \citealt{Motta2020, Pintore2021}) and flux-dips in soft ULXs (e.g. \citealt{Stobbart2004,Feng2016,Alston2021,Pinto2021, D'Ai2021}). Pointed, deep observations with high-throughput telescopes are limited and cannot enable a detailed study of the long-term behaviour. This prevents us from obtaining an overall view of their accretion cycle. \textit{Swift} instead allows us to perform X-ray spectroscopic studies over a long ($\sim$ yearly) baseline with weekly visits, as proved by previous studies  \citep[e.g.][]{Kaaret2009,Grise2010,LaParola2015, Brightman2020,Gurpide2021b}.

\begin{figure*}
\begin{center}

\includegraphics[width=16.0cm]{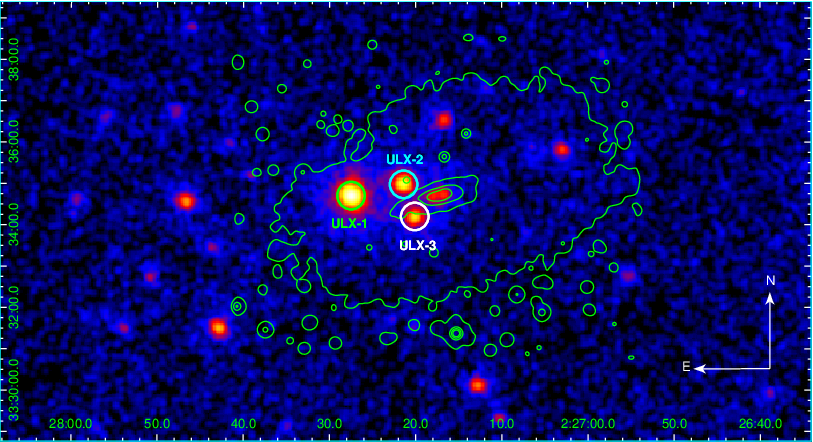}

   \caption{Stacked image from all the \swift\ observations analysed in this work. The coordinates of the image are expressed in RA, Dec (FK5). Positions of ULX-1, ULX-2 and ULX-3 are indicated by a green, cyan and white circle of 20$^{\prime\prime}$ radius, respectively. The optical contours of the galaxy NGC 925 from DSS are super-imposed with solid green lines for reference.}
   \label{Fig:xray_img}
   \end{center}
\end{figure*}

The long-term behaviour is reported only for a small fraction of the whole ULX population (e.g. \citealt{Brightman2019,Brightman2020,Gurpide2021b}), containing 1843 ULX candidates according to the recent catalogue of \citealt{Walton2021}. Thus, to enlarge this sample, it is crucial to perform studies of long-term variability, possibly correlated with spectral changes, and to look for periodicities. 
The main aim of this work is to report on the long-term properties of the ULXs in the galaxy NGC 925 (Figure \ref{Fig:xray_img}),
using data taken with a \swift\ monitoring. To complete our analysis we used also archival data taken with other X-ray facilities.

\begin{table}
\begin{center} 
\begin{tabular}{ccccc} \hline
 & RA & Dec & Chandra & ref.\\
\hline
ULX-1 & 02:27:27.5 & +33:34:43.0 & CXO  J022727+333443 & a\\
ULX-2 & 02:27:21.5 & +33:35:00.7 & CXO J022721+333500 & a\\
ULX-3 & 02:27:20.2 & +33:34:12.8 & & b\\
\hline
\end{tabular}
\end{center}
\label{tab:coord}
\caption{Chandra coordinates of the three ULXs in the galaxy NGC 925. Column 4 lists the names of the sources detected in the Chandra catalog, while column 5 reports the first reference for each source: a) \citealt{swartz2011}; b) \citealt{Earnshaw2020}.}
\end{table}

NGC 925 is a spiral galaxy (SAB(s)d) at a distance of $\sim$ 8.9 Mpc \citep{Tully2013}. It contains 3 known ULXs:  NGC 925 ULX-1,  
NGC 925 ULX-2,   
whose data were extensively studied in \citet{Pintore2018b}, and NGC 925 ULX-3, which is a transient ULX discovered by \citet{Earnshaw2020}. Their positions as measured by Chandra (\citealt{swartz2011,Earnshaw2020}) are reported in Table~\ref{tab:coord}.

The paper is organized as follows. In Sect. \ref{sec:data_red}, we describe the data used in this paper, the basic data reduction and analysis. The results for each ULX in the galaxy are reported in Sect. \ref{sec:results}, while their implications are discussed in Sect. \ref{sec:discussion}. Finally our conclusions are in Sect. \ref{sec:conclusion}.

\begin{table}
\scalebox{0.84}{\begin{minipage}{24cm}
\begin{tabular}{@{\makebox[1.3em][l]{\rownumber\space}} | llccr}
\hline
Instr. & Obs.ID & Start time & Stop time & Exp \\
 & & \multicolumn{2}{c}{[YYYY-MM-DD hh:mm:ss]} & [ks] 
\gdef\rownumber{\stepcounter{magicrownumbers}\arabic{magicrownumbers}} \\
\hline
\swift & 00045596001 & 2011-07-21T09:26:28 & 2011-07-21T15:59:55 & 2.6\\
\swift & 00045596002 & 2011-07-26T14:40:42 & 2011-07-26T14:54:55 & 0.8\\
\swift & 00045596003 & 2011-07-27T00:03:18 & 2011-07-27T14:58:52 & 5.7\\
\swift & 00045596004 & 2011-07-31T02:25:56 & 2011-07-31T23:19:55 & 0.3\\
\swift & 00045596005 & 2011-08-02T00:58:52 & 2011-08-02T02:36:56 & 0.2\\
\swift & 00045596006 & 2011-08-06T15:39:11 & 2011-08-06T15:43:56 & 0.3\\
\swift & 00045596007 & 2011-09-07T03:33:50 & 2011-09-07T03:55:57 & 1.3\\
\swift & 00045596008 & 2011-09-13T00:48:18 & 2011-09-13T12:09:56 & 3.6\\
\swift & 00045596009 & 2012-08-23T22:06:46 & 2012-08-23T23:45:54 & 0.6\\
\swift & 00045596010 & 2012-08-24T17:00:02 & 2012-08-25T14:00:55 & 5.3\\
\swift & 00045596011 & 2012-08-27T20:28:20 & 2012-08-27T20:43:55 & 0.9\\
\swift & 00045596012 & 2012-08-28T02:54:05 & 2012-08-28T20:28:55 & 4.4\\
\swift & 00045596013 & 2012-08-29T01:16:16 & 2012-08-29T22:26:56 & 6.5\\
\swift & 00045596014 & 2014-09-07T07:31:45 & 2014-09-07T09:19:54 & 2.1\\
\swift & 00045596015 & 2017-11-19T05:59:52 & 2017-11-19T07:47:53 & 2.1\\
\swift & 00045596016 & 2017-11-19T15:56:16 & 2017-11-19T17:27:52 & 1.9\\
\swift & 00045596017 & 2017-11-21T07:45:23 & 2017-11-21T09:30:52 & 1.7\\
\swift & 00045596018 & 2017-11-25T20:05:50 & 2017-11-25T22:00:54 & 1.8\\
\swift & 00045596019 & 2019-08-18T16:33:23 & 2019-08-18T22:55:52 & 1.3\\
\swift & 00045596020 & 2019-08-21T04:41:38 & 2019-08-21T05:02:52 & 1.3\\
\swift & 00045596021 & 2019-08-25T18:33:26 & 2019-08-25T18:50:51 & 1.0\\
\swift & 00045596022 & 2019-08-27T05:51:49 & 2019-08-27T21:57:53 & 1.5\\
\swift & 00045596023 & 2019-09-01T05:20:20 & 2019-09-01T16:49:53 & 1.6\\
\swift & 00045596024 & 2019-09-08T06:00:36 & 2019-09-09T03:01:52 & 2.4\\
\swift & 00045596025 & 2019-09-15T13:27:33 & 2019-09-15T23:14:53 & 2.6\\
\swift & 00045596026 & 2019-09-22T22:23:05 & 2019-09-22T22:47:52 & 1.5\\
\swift & 00045596027 & 2019-09-25T22:05:20 & 2019-09-25T22:28:54 & 1.4\\
\swift & 00045596028 & 2019-10-06T02:10:34 & 2019-10-06T21:28:54 & 1.2\\
\swift & 00045596029 & 2019-10-09T11:29:11 & 2019-10-09T19:31:51 & 1.6\\
\swift & 00045596030 & 2019-10-13T15:26:57 & 2019-10-13T17:28:52 & 2.9\\
\swift & 00045596031 & 2019-10-20T06:55:11 & 2019-10-20T21:41:52 & 2.1\\
\swift & 00045596032 & 2019-10-27T00:03:15 & 2019-10-27T22:35:52 & 2.5\\
\swift & 00045596033 & 2019-11-03T18:17:09 & 2019-11-03T21:53:52 & 2.8\\
\swift & 00045596034 & 2019-11-13T12:48:09 & 2019-11-13T22:33:53 & 2.8\\
\swift & 00045596035 & 2019-11-17T06:02:31 & 2019-11-18T10:54:53 & 2.8\\
\swift & 00045596036 & 2019-11-28T19:07:33 & 2019-11-28T22:43:54 & 2.5\\
\swift & 00045596037 & 2019-12-08T03:37:33 & 2019-12-08T15:12:53 & 3.3\\
\swift & 00089002001 & 2019-12-13T03:19:50 & 2019-12-13T05:02:54 & 1.9\\
\swift & 00045596038 & 2019-12-18T12:12:51 & 2019-12-18T15:39:53 & 1.6\\
\swift & 00045596039 & 2019-12-22T19:45:59 & 2019-12-22T19:53:53 & 0.5\\
\swift & 00045596041 & 2020-01-02T09:15:03 & 2020-01-02T14:22:52 & 3.5\\
\swift & 00045596042 & 2020-01-07T18:45:42 & 2020-01-07T22:06:52 & 1.2\\
\swift & 00045596044 & 2020-03-01T21:11:49 & 2020-03-01T23:08:53 & 2.4\\
\swift & 00045596045 & 2020-03-08T09:21:10 & 2020-03-08T11:21:09 & 2.4\\
\swift & 00045596046 & 2020-03-15T00:37:12 & 2020-03-15T03:54:52 & 2.9\\
\swift & 00095702001 & 2020-07-01T01:01:15 & 2020-07-01T09:06:53 & 2.1\\
\swift & 00095702002 & 2020-07-11T04:45:24 & 2020-07-11T06:35:54 & 2.6\\
\swift & 00095702003 & 2020-07-21T04:02:28 & 2020-07-21T13:45:53 & 1.9\\
\swift & 00095702004 & 2020-07-31T17:17:45 & 2020-07-31T22:10:53 & 2.0\\
\swift & 00095702005 & 2020-08-10T08:14:50 & 2020-08-10T10:28:53 & 2.6\\
\swift & 00089004001 & 2020-08-17T07:32:05 & 2020-08-17T07:58:52 & 1.6\\
\swift & 00095702006 & 2020-08-20T00:52:59 & 2020-08-20T16:55:52 & 2.8\\
\swift & 00095702007 & 2020-08-30T17:18:07 & 2020-08-30T19:04:53 & 2.3\\
\swift & 00095702008 & 2020-09-09T09:58:56 & 2020-09-09T12:04:52 & 2.5\\
\swift & 00095702009 & 2020-09-19T10:41:53 & 2020-09-19T14:05:54 & 2.4\\
\swift & 00095702010 & 2020-09-29T19:24:34 & 2020-09-29T21:16:52 & 2.5\\
\swift & 00095702011 & 2020-10-08T13:35:29 & 2020-10-09T21:33:53 & 2.6\\
\swift & 00095702012 & 2020-10-19T12:46:33 & 2020-10-19T12:48:54 & 0.1\\
\swift & 00095702013 & 2020-10-22T12:09:32 & 2020-10-22T13:53:54 & 2.3\\
\swift & 00095702014 & 2020-10-29T14:49:06 & 2020-10-29T16:32:52 & 1.9\\
\swift & 00095702015 & 2020-11-08T10:40:54 & 2020-11-08T12:28:53 & 2.5\\
\swift & 00095702016 & 2020-11-18T16:12:05 & 2020-11-18T22:50:54 & 1.5\\
\swift & 00095702017 & 2020-11-29T13:15:24 & 2020-11-29T16:52:53 & 2.2\\
\swift & 00095702018 & 2020-12-08T05:51:23 & 2020-12-08T07:48:53 & 2.0\\
\swift & 00095702019 & 2020-12-18T11:15:01 & 2020-12-18T17:53:53 & 1.9\\
\swift & 00095702020 & 2020-12-28T10:18:55 & 2020-12-28T10:20:26 & 0.1\\
\swift & 00095702021 & 2021-01-01T06:37:39 & 2021-01-01T08:25:52 & 2.4\\
\swift & 00095702022 & 2021-01-07T12:31:39 & 2021-01-07T16:08:52 & 2.5\\
\swift & 00095702023 & 2021-01-17T02:01:45 & 2021-01-17T08:35:54 & 2.5\\
\swift & 00095702025 & 2021-01-31T10:15:05 & 2021-01-31T13:37:53 & 2.3\\
\swift & 00095702026 & 2021-02-06T13:58:46 & 2021-02-06T15:47:53 & 2.5\\
\swift & 00095702027 & 2021-02-15T00:28:25 & 2021-02-16T13:37:54 & 2.6\\
\swift & 00095702028 & 2021-02-26T01:04:10 & 2021-02-26T22:07:53 & 2.2\\
\swift & 00095702029 & 2021-03-08T12:36:39 & 2021-03-08T14:33:53 & 2.4\\
\chandra & 7104 & 2005-11-23T07:53:09 & 2005-11-23T08:57:05 & 2.2\\
\chandra & 20356 & 2017-12-01T03:38:39 & 2017-12-01T07:04:06 & 10.0\\
\textit{XMM-N.} & 0784510301 & 2017-01-18T19:45:14 & 2017-01-19T09:38:34 & 50.0\\
\nus & 30201003002 & 2017-01-18T19:36:09 & 2017-01-19T17:41:09 & 42.6\\
\nus & 90501351002 & 2019-12-12T05:31:09 & 2019-12-13T11:46:09 & 53.0\\
\hline
\end{tabular}
\end{minipage}}

\caption{Log of the observations used in this work.}
\label{log}
\end{table}

\section{Data Reduction and Analysis}
\label{sec:data_red}
\subsection{Swift/XRT} 
\swift\ pointed at the galaxy NGC 925 for 74 times between July 2011 and March 2021 (see the stacked image of all the pointings in Figure \ref{Fig:xray_img}), with observations in PC mode of, on average, about 2 ksec. We list them, with dates of observation and relative exposures, in Tab.~\ref{log}. 

We reduced \swift\ data adopting standard procedures\footnote{\url{https://www.swift.ac.uk/analysis/xrt/index.php}}. We ran the \textsc{xrtpipeline}\footnote{\url{https://heasarc.gsfc.nasa.gov/ftools/caldb/help/xrtpipeline.html}}  
and we used a sliding-cell method of detection (\textsc{detect} command\footnote{\url{https://heasarc.gsfc.nasa.gov/docs/xanadu/ximage/manual/node7.html}} in \textsc{ximage}\footnote{\url{https://heasarc.gsfc.nasa.gov/docs/xanadu/ximage/ximage.html}}), to verify if new ULXs turned on during the \swift\ observations. We choose a detection threshold of 3$\sigma$ for each observation and for the stacked set of all observations listed in Table~\ref{sec:data_red}. 
However, no new ULXs were found in the D$_{25}$ of the galaxy (D$_{25}\sim$ 10.7$'$ from HyperLeda database)\footnote{\url{http://leda.univ-lyon1.fr}; \citealt{Makarov2014}}.

To obtain light-curves and spectra of all the ULXs in the galaxy, we used the online tool for XRT products\footnote{\url{https://www.swift.ac.uk/user_objects/}} \citep{evans09}. The extraction regions have been centred on the coordinates obtained from the detection on the \swift\ stacked image, which are fully consistent with the \chandra\ coordinates. 
Since the minimum searching radius allowed for the centroiding (1 arcmin) is comparable with the distances among the ULXs in NGC 925, the centroiding option has been disabled, in order to avoid the possible spurious centering on the wrong ULX. 
We also created an average \swift\ spectrum for each ULX, from the stacked image.

\subsection{Chandra} 
\chandra/ACIS-S observed the galaxy twice on 25 November 2006 and 1st December 2018 (Obs.ID: 7104, 20356) with observing time of 2 and 10 ksec, respectively (Tab.~\ref{log}).

We reduced Chandra data with {\sc CIAO} v.4.11 \citep{Fruscione2006} and used calibration files CALDB v.4.8.3. We selected source events from a circular region of 3$^{\prime\prime}$ radius (the smallest suitable radius owing to the off-axis position of the sources, especially in the first observation), while the background was chosen from annular regions with inner and outer radii of 6$^{\prime\prime}$ and 20$^{\prime\prime}$, respectively. We extracted  sources and background spectra, and their response and ancillary files using the {\sc CIAO} task \textsc{specextract}.

\subsection{XMM-Newton} 
There is only one public \xmm\ observation taken on 18 January 2017 at the time of the analysis (Tab.~\ref{log}). We reduced the EPIC-pn and MOS data with the software SAS v17.0.0. We selected single and double pixel events  for the pn (PATTERN $\leq$ 4) and single and multiple pixel events for the MOS (PATTERN $\leq$ 12). We cleaned the data to remove time intervals of high particle background, which resulted in a net exposure time of $\sim$ 32 ks (pn) and 42 ks (MOS). 

We extracted source data from circular regions of 35$^{\prime\prime}$ (ULX-1) and 20$^{\prime\prime}$ (ULX-2, ULX-3) radii. We used a smaller extraction region for ULX-2 and ULX-3 to prevent contamination between the two sources, which are close to each other. ULX-2 falls into a CCD gap in the EPIC-pn, so we did not use the pn data for the spectral analysis of ULX-2. 
The background data were extracted from circular regions of 60$^{\prime\prime}$ radius, in a region free of contaminating sources, in the same CCD quadrant and near each source, so as to be at a similar off-axis angular distance.

\subsection{NuSTAR}
Two \nus\ observations were taken in 2017 ($\sim$ 42 ks) and 2019 ($\sim$ 53 ks). We note that the first one started about 20 minutes before the 2017 \xmm\ observation (Tab.~\ref{log}).

\nus\ data were reduced using standard procedures with {\sc nustardas} v1.3.0  
of the Ftools v6.16 and we used the CALDB v.20180312. The extraction region for ULX-1 was a circle of 50$^{\prime\prime}$ radius. The background was chosen from a nearby 80$^{\prime\prime}$ radius circular region, and free of sources. 
For ULX-2 we used a circular region of 30$^{\prime\prime}$ to limit any possible contamination from the nearby ULX-1 and ULX-3. 

In 2017 only ULX-1 and ULX-2 were detected; we used the \nus\ data in conjunction with simultaneous \xmm\ data. In 2019, instead, also ULX-3 was bright, and the closeness to ULX-2 does not allow us to correctly disentangle the two. Therefore we used only the \nus\ observation of ULX-1 in conjunction with three \swift\ observations close in time (8, 13 and 18 December 2019). We combined the latter to get a stacked, quasi-simultaneous ULX-1 spectrum to be analysed in conjunction with the \nus\ spectrum.
\newline
\newline
\newline

\subsection{Spectral analysis} 

\begin{table*}
\begin{center} 

\scalebox{0.9}{%
\begin{tabular}{l|l|cccccccccc} 
\hline

 Source & Epoch & n$_{\rm H}^{int}$ & kT$_\text{diskbb}$  &  N$_\text{disc}$  & kT$_\text{bbodyrad}$ & N$_\text{bb}$ 
 & Flux$_\text{diskbb}$ & Flux$_\text{bbodyrad}$ 
 & $\chi^{2}$/dof & P$_{val}$\\
 & & $10^{22}$ cm$^{-2}$    &  keV   &   10$^{-3}$    & keV       &  &
 \multicolumn{2}{c}{$10^{-13}$ erg s$^{-1}$ cm$^{-2}$} & \\ 
\hline
ULX1 &  &  & & &  & &  & \\
\hline
& XMM & 0.09$^{+0.02}_{-0.02}$ & 2.3$^{+0.1}_{-0.1}$ & 3.7$^{+0.8}_{-0.7}$ & 0.29$^{+0.02}_{-0.02}$ & 5.6$^{+1.7}_{-1.2}$ 
& 18.5$\pm$0.4  & 4.1$\pm$0.1 
&1153.29/1131 & 0.32\\
& XMM/NuSTAR$^{(a)}$ & 0.06$^{+0.02}_{-0.02}$ & 2.8$^{+0.1}_{-0.1}$ & 1.7$^{+0.3}_{-0.3}$ & 0.33$^{+0.02}_{-0.02}$ & 3.7$^{+0.8}_{-0.6}$ 
& 18.0$\pm$0.4 & 4.7$\pm$0.1 
& 1371.67/1269 & 0.02\\
& Chandra & 0.09 & 2.2$^{+0.8}_{-0.4}$ & 4.2$^{+4.7}_{-2.8}$ & 0.3$^{+0.1}_{-0.1}$ & 2.9$^{+3.7}_{-1.6}$ 
& 19.0$\pm$2.0 / 18.8$\pm$1.1 & 2.3$\pm$0.7 / 2.5$\pm$0.6 
& 70.89/63 & 0.29\\
& Swift average & 0.09 & 2.1$^{+0.3}_{-0.2}$ & 4.7$^{+2.1}_{-1.6}$ & 0.3$^{+0.1}_{-0.1}$ & 2.7$^{+2.6}_{-1.2}$ 
 & 17.4$\pm$0.5 & 1.8$\pm$0.2 
 & 181.47/206 & 0.81\\
& Swift/NuSTAR$^{(b)}$ & 0.09 & 3.7$^{+0.5}_{-0.4}$ & 0.2$^{+0.1}_{-0.1}$ & 0.4$^{+0.1}_{-0.1}$ & 0.9$^{+1.0}_{-0.6}$ 
 & 5.3$\pm$2.1 & 3.5$\pm$0.8  
 & 87.18/79 & 0.25\\
\hline
ULX2 &  &  & & &  & &  & \\
\hline

  &  XMM/NuSTAR & 0.02$^{+0.02}_{-0.02}$ & 2.9$^{+0.7}_{-0.5}$ & 0.2$^{+0.2}_{-0.1}$ & 0.31$^{+0.04}_{-0.04}$ & 1.1$^{+1.0}_{-0.5}$  
  & 2.6$\pm$0.2 & 1.1$\pm$0.1 & 145.27/139 & 0.34\\
 &  Chandra & 0.02 & 1.5$^{+3.9}_{-0.5}$ & 3.1$^{+10.1}_{-3.0}$ & 0.2$^{+0.1}_{-0.1}$ & 3.3$^{+12.6}_{-2.7}$ 
 & 3.2$\pm$0.9 / 3.1$\pm$0.4 & 0.8$\pm$0.6 / 1.0$\pm$0.3 & 4.23/9 & 0.90\\
 &  Swift average & 0.02 & 1.3$^{+0.1}_{-0.1}$ & 7.3$^{+2.0}_{-1.6}$ & 0.03$^{+0.04}_{-0.03}$ & $>$242.5 
 & 4.2$\pm$0.2 & 0.5$\pm$0.3 & 74.59/62 & 0.13\\
 \hline
ULX3 &  &  & & &  & &  & \\
\hline
 & XMM & 0.22$^{+0.04}_{-0.03}$ & 1.1$^{+0.6}_{-0.3}$ & 0.6$^{+1.8}_{-0.5}$ & 0.12$^{+0.02}_{-0.02}$ & 17.9$^{+24.4}_{-9.9}$
 & 0.26$\pm$0.04 & 0.3$\pm$0.1 & 40.55/38 & 0.36\\
  & Chandra & 0.22 & 1.6$^{+0.4}_{-0.3}$ & 4.0$^{+3.1}_{-2.4}$ & 0.16$^{+0.05}_{-0.04}$ & 49.0$^{+120.9}_{-37.4}$ 
  & -- / 5.2$\pm$0.5 & -- / 2.6$\pm$0.7 &  17.98/18 & 0.46\\
 & Swift average & 0.22 & 1.5$^{+0.3}_{-0.2}$ & 2.2$^{+1.6}_{-1.0}$ & 0.12$^{+0.03}_{-0.02}$ & 65.3$^{+158.4}_{-43.4}$ 
 & 1.9$\pm$0.2 & 1.2$\pm$0.2 & 34.43/32 & 0.35\\
\hline
\end{tabular}
}
\end{center}
\caption{Spectral parameters for the different spectra of  ULX-1, ULX-2, ULX-3: \xmm, \xmm\ / \textit{NuSTAR}, \chandra\ and the \swift\ average spectra. The model used is {\sc const  $\times$ tbabs $\times$ (bbodyrad + diskbb)}, where the n$_{H}^{int}$ reported in table is the intrinsic n$_{H}$, while the foreground Galactic absorption column is fixed to n$_{H}^{gal} = 7.26\times$10$^{20}$ cm$^{-2}$. Errors are computed at the 90\% confidence level for each parameter of interest.
Unabsorbed  X-ray fluxes, derived by using the pseudo-model {\sc cflux} in {\sc xspec}, are calculated in the 0.3--10 keV energy band in units of 10$^{-13}$ erg s$^{-1}$ cm$^{-2}$; the \xmm\ fluxes are from EPIC-pn for ULX-1 and ULX-3, and from MOS1 for ULX-2. The two \chandra\ observations are fit together: the flux on the left/right refers to observation 7104/20356. The n$_{\rm H}$ is fixed to the value obtained with \xmm\ (ULX-1 and ULX-3) or \xmm\ / \nus\ fit (ULX-2).\\
In the last column we report the null hypothesis probability P$_{val}$, i.e. the probability that the observed data have been drawn from the model, given the $\chi^{2}$ value and the degrees of freedom. We consider acceptable all the fit with P$_{val} \geq 0.05$ that correspond to a confidence level larger than 2$\sigma$.\\
$^{(a)}$ The fit is not statistically acceptable, since the model leaves skewed residuals above 10 keV,  suggesting the need for a third spectral component, see details in the text (section \ref{sec:results_ULX-1} and Figure \ref{Fig:residui_mos_ulx2}).\\
$^{(b)}$ For ULX-1 we also fit the \swift\ (8, 13, 18 December 2019 average) together with the \nus\ 2019 spectrum.}

\label{tab:specULX}
\end{table*}

\begin{figure}
 \centerline{\includegraphics[width=.65\columnwidth, angle=270]{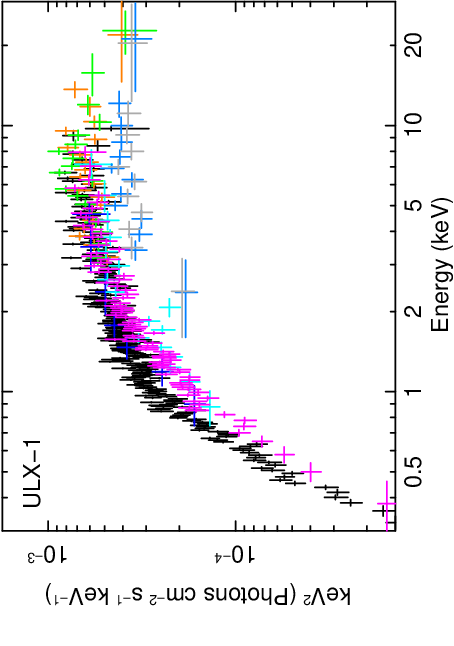}}
 \centerline{\includegraphics[width=.65\columnwidth, angle=270]{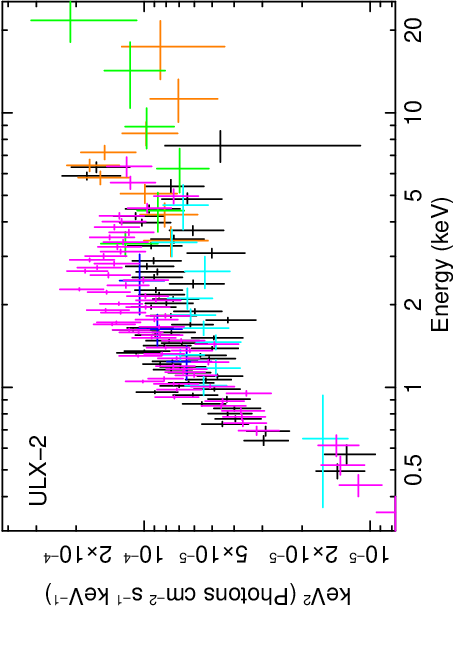}}
 \centerline{\includegraphics[width=.65\columnwidth, angle=270]{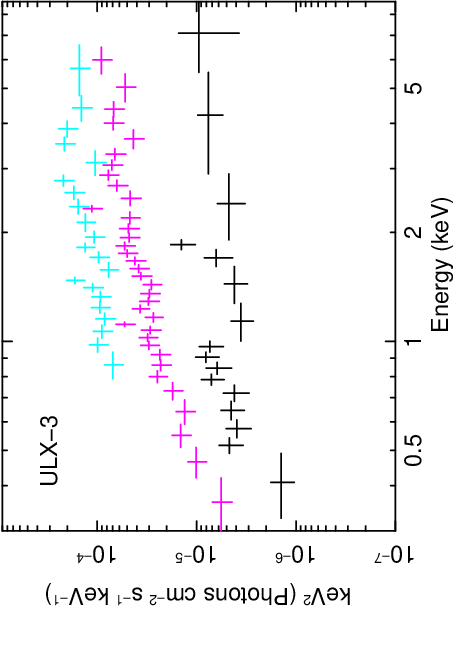}}
   \caption{Unfolded spectra ($E^2 f(E)$) for the three ULXs (data rebinned for display purposes) using a powerlaw model with photon index $\Gamma=0$. \xmm\ EPIC-pn data are in black (for ULX-2 we plot instead the MOS1 data); NuSTAR FPMA 2017 (orange), NuSTAR FMPB 2017 (green); NuSTAR FPMA 2019 (grey); NuSTAR FMPB 2019 (light-blue); Chandra 7104 (blue); Chandra 20356 (cyan); Swift/XRT average (magenta). \\ Upper panel: ULX-1; Central panel: ULX-2; Bottom panel: ULX-3.
   }
   \label{Fig:unfold_spec}
\end{figure}

We carried out the spectral analysis by using {\sc xspec} v.12.10.1 \citep{Arnaud1996}. We grouped the spectra from all the instruments with at least 20 counts per energy bin and we applied the $\chi^{2}$ statistics in the spectral analysis.
In all the spectral fits, we included an absorbtion component modeled with {\sc tbabs} and using the abundances of \citet{lodders2003}.

Different models have been used in the literature to fit ULXs spectra, such as power-law, disc-like and comptonization continua \citep[see e.g.][]{Walton2013, Brightman2016, pintore2016, Bachetti2013, Middleton2011, Roberts2006, Fuerst2017}. 

ULXs spectra with low statistics (e.g. those obtained with \swift) can usually be fitted with simple models with one component. For high statistics data (e.g. \xmm) instead, at least two components are needed to explain the complex spectral shape of ULXs \citep[e.g.][]{Gladstone2009b, Pintore2011, Sutton2013, middleton2015a, Walton2020}. 
The need of two components models can be explained if we assume a regime of super-Eddington accretion onto a stellar mass compact object, where the disc is expected to be puffed up by the huge radiation pressure inside the spherization radius (see, e.g., \citealt{Poutanen2007}). A complex structure would likely arise implying the formation of zones with different temperatures and dominant radiation processes. In high counts X-ray spectra of ULXs, multiple components are often required in order to reproduce a hot, harder, inner accretion flow and a cooler, softer, outer disc component (see, e.g., \citealt{Sutton2013}). The hard component is commonly very broad with a strong turnover above 5 keV, while the soft component can be generally fitted with a simple black body or multi-colour black body disc. The latter component is usually referred as the outer disc and/or the photosphere of the optically thick outflows (see, e.g., \citealt{Stobbart2006,Gladstone2009b,Walton2020,Gurpide2021a}). 

Motivated by the above-mentioned considerations on ULXs spectral-fitting, we first analysed the highest quality data, \xmm\ + simultaneous (2017) \nus\ data, for ULX-1 and ULX-2 and just \xmm\ data for ULX-3, with two-component models. 
A good fit resulted from a two thermal components model for ULX-2 and ULX-3, i.e. a black body ({\sc bbodyrad} in {\sc xspec}) plus a multi-colour disc ({\sc diskbb}; \citealt{Mitsuda1984}), while for ULX-1 a third harder spectral component, {\sc cutoffpl} in {\sc xspec}, was identified in the spectrum. The blackbody models the softer part of the spectra while the disc blackbody the harder part. We added also a multiplicative constant, \textsc{const} in {\sc xspec},  to take into account different flux calibrations amongst instruments: all constants are consistent within 10\%, as expected from e.g. \citet{Madsen2017}. 
  
For consistency, we applied the {\sc bbodyrad} + {\sc diskbb} model also to poorer statistics \chandra\ data and to the average \swift\ spectra. The absorbing column, composed of a fixed Galactic absorption (n$_{H}^{gal}$ = 7.26$\times$10$^{20}$ cm$^{-2}$)\footnote{\url{https://heasarc.gsfc.nasa.gov/cgi-bin/Tools/w3nh/w3nh.pl}} plus an intrinsic absorption component, was not always well constrained, so we decided to fix it at the best value obtained with the \xmm/\nus\ (for ULX-2) data or \xmm\ (for ULX-1, because the fit of the \xmm\ / \nus\ spectrum with two components was not statistically acceptable and in the fit with three components the n$_{\rm H}$ was not well constrained, and ULX-3, which was not detected in the \nus\ data).  
The best-fitting spectral parameters are reported in Table~\ref{tab:specULX}. The unfolded spectra of each ULX are reported in Figure~\ref{Fig:unfold_spec}. 

\subsection{Hardness Ratio}
\begin{figure}
 \centerline{\includegraphics[width=1.1\columnwidth, angle=0]{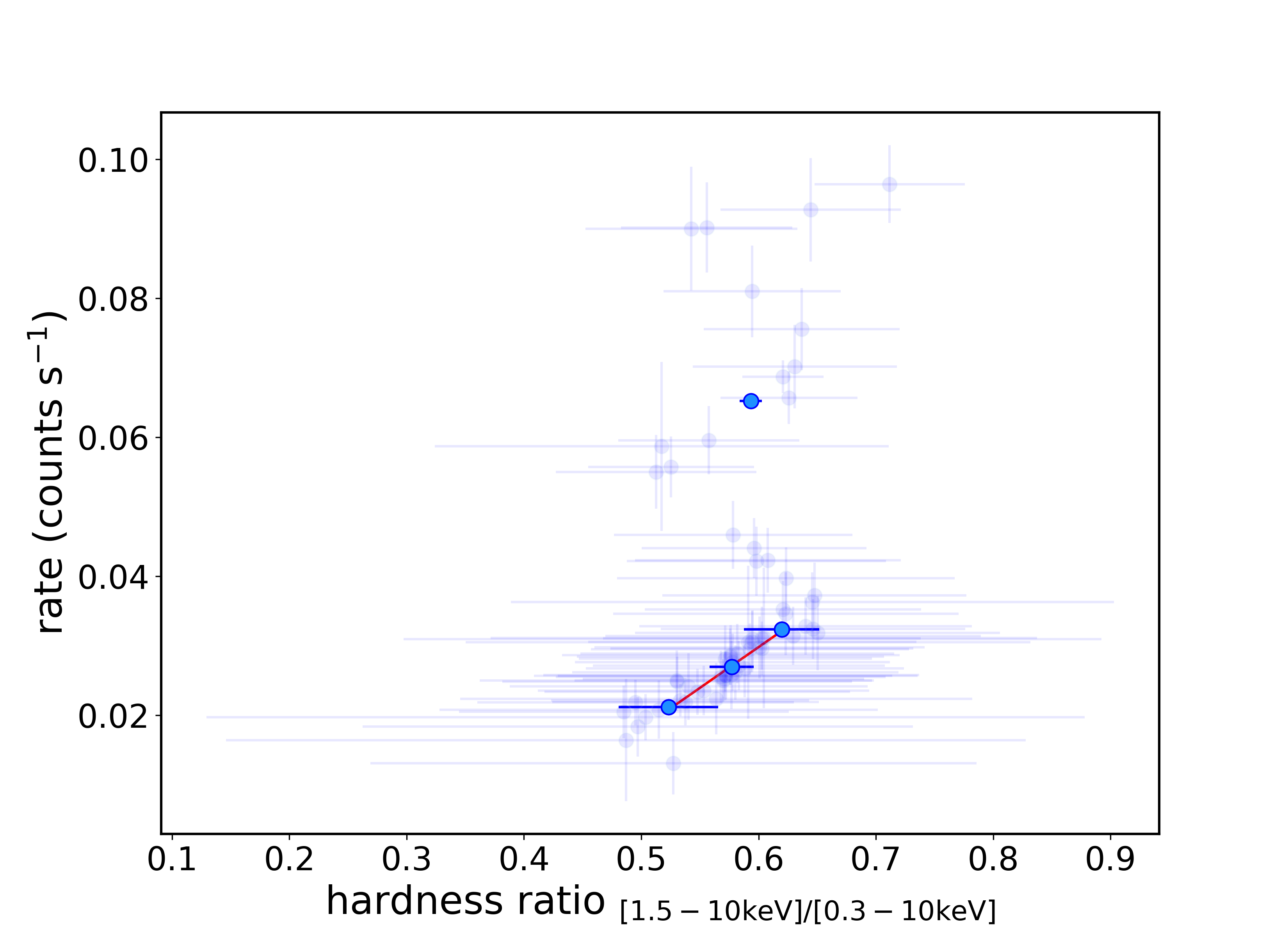}}
 \centerline{\includegraphics[width=1.1\columnwidth, angle=0]{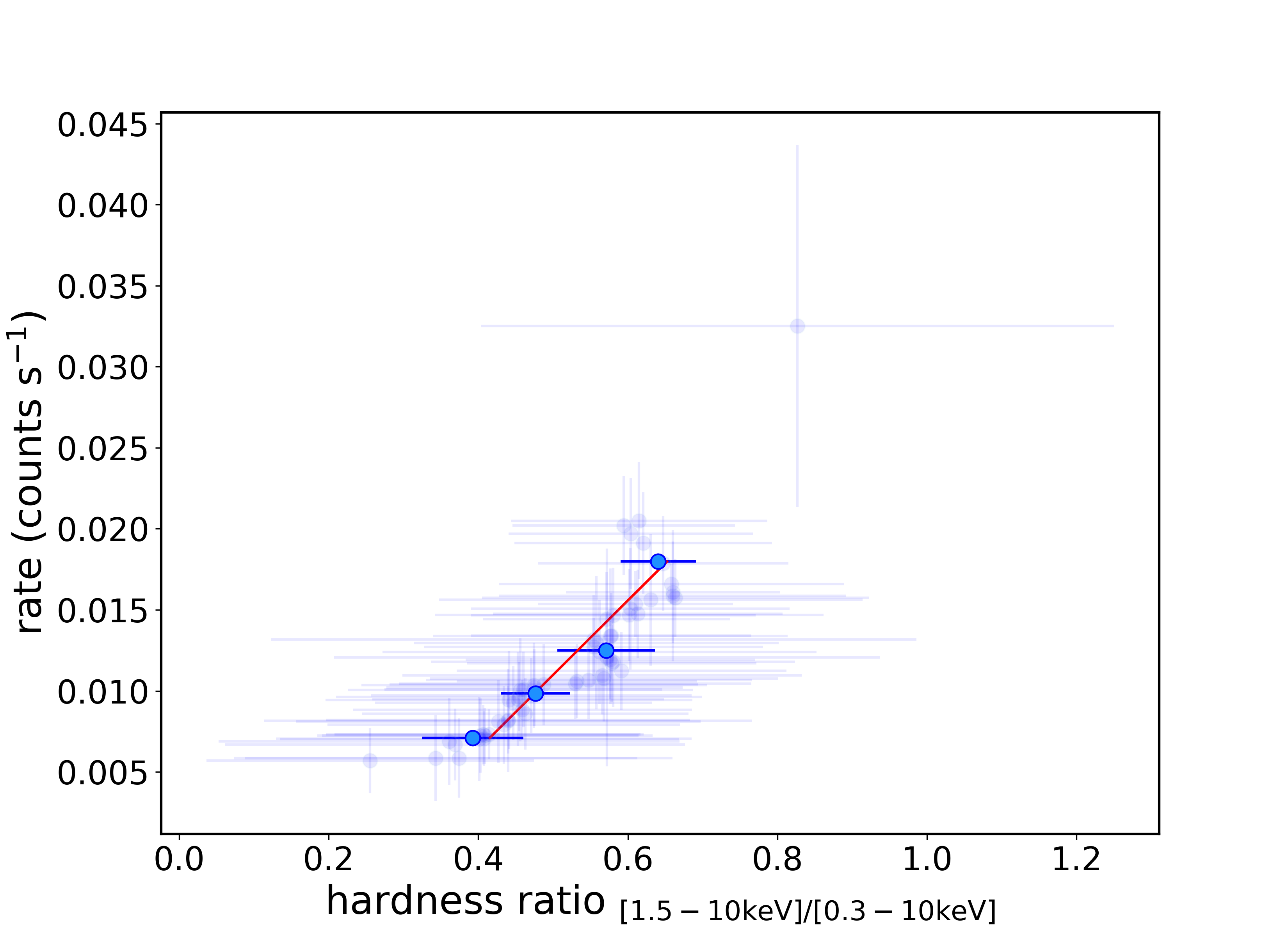}}
 \centerline{\includegraphics[width=1.1\columnwidth, angle=0]{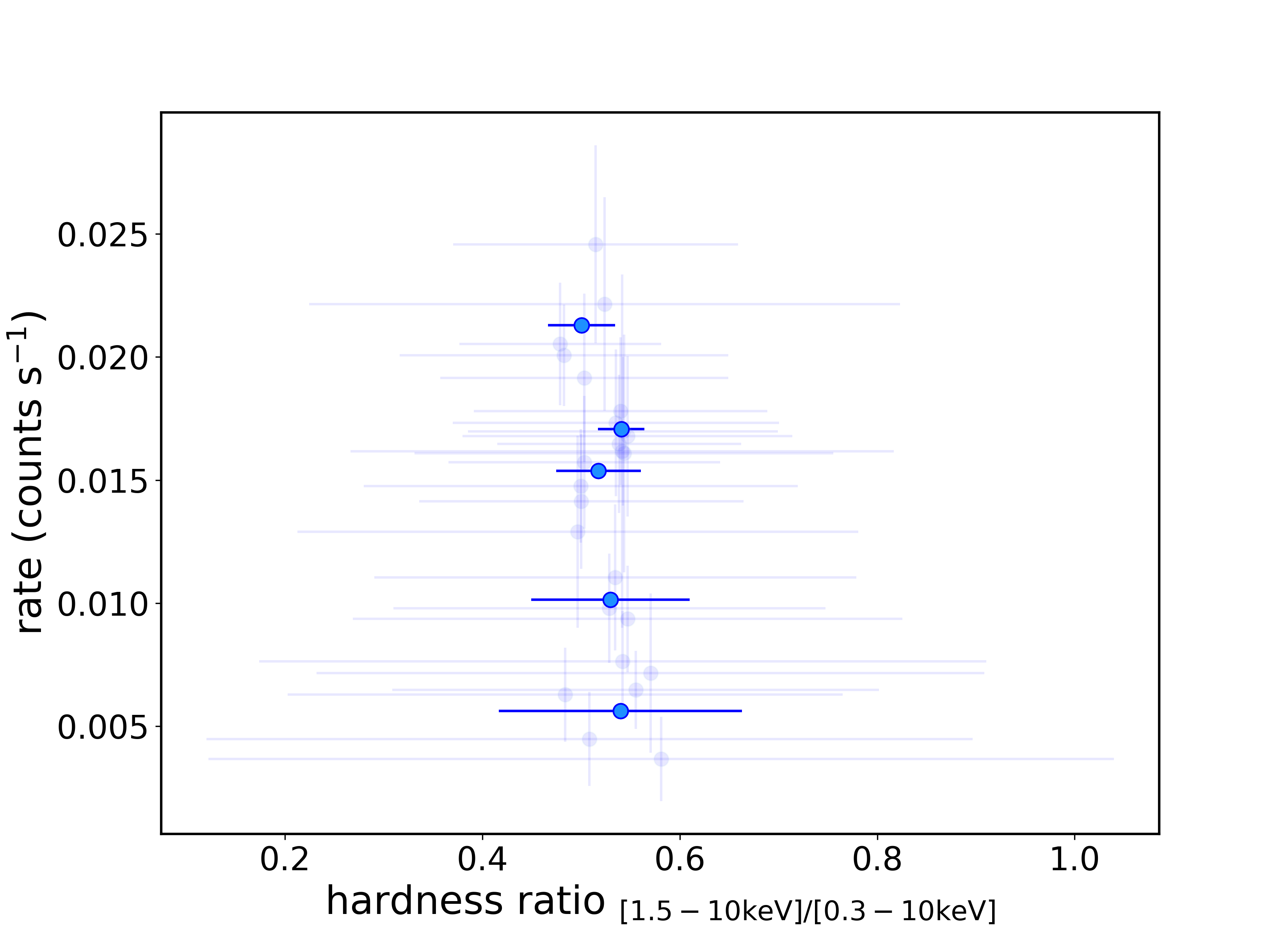}}
   \caption{\swift\ count rate vs. hardness ratio for ULX-1 (upper); ULX-2 (middle); ULX-3 (bottom). The count rate is calculated in the (0.3-10 keV) band, in units of counts s$^{-1}$. The hardness ratio is the ratio between the count rates in (1.5-10) and (0.3-10) keV. The solid blue dots are the binned hardness ratios as a function of the count rate, while the transparent blue dots are the unbinned hardness ratios. The super-imposed solid line is the result of the linear regression applied to the binned points for: the hardness at low count rates ($<$ 0.04 cts s$^{-1}$) for ULX-1 (upper); all the binned points for ULX-2 (middle).}
  \label{Fig:hardness_hard_su_tot}
\end{figure}
We derived a hardness--intensity diagram (HID) for each \swift\ observation, where the hardness was calculated as the ratio between the count rates in the hard (1.5-10 keV) and total (0.3-10 keV) energy bands, in each observation. We chose 1.5 keV as the threshold since it is about the median energy where the {\sc bbodyrad} and {\sc diskbb} spectral components cross for the three ULXs. 
Slight differences in this pivot energy are seen for the three sources, but we decided to adopt the same value for all the sources in order to allow a direct comparison of the results. 
The error bars of each point in the HID are large and did not permit us to identify clearly any trend or spectral evolution with the count rate. Therefore, in order to reduce data noise, we binned the hardness ratio based on the count rate. In particular, we grouped the data in order to have at least 15 observations in any given count rate bin for ULX-1 and ULX-2. Instead, for ULX-3, because of the lower statistics, we constructed bins containing at least 5 observations (see Figure ~\ref{Fig:hardness_hard_su_tot}). 

From the HID, we note a possible linear trend for ULX-1 and ULX-2 at low count rates. We applied a linear fit to the binned data to verify it; we performed the fit applying the minimum least squares method. We report the results in sections \ref{sec:results_ULX-1} and \ref{sec:results_ULX-2}.
\newline
\newline

\subsection{Timing analysis}
\label{sec:timing-analysis}
\begin{table}
\begin{center} 
\begin{tabular}{cccc} \hline
  & soft$_{[0.3-1.5] keV}$ & hard$_{[1.5-10.0] keV}$ & total$_{[0.3-10.0] keV}$ \\
\hline
ULX-1 & 0.54$\pm$0.01 & 0.59$\pm$0.01 & 0.54$\pm$0.01\\
ULX-2 & 0.26$\pm$0.05 & 0.59$\pm$0.02 & 0.35$\pm$0.02\\
ULX-3 & 0.57$\pm$0.01 & 0.38$\pm$0.01 & 0.63$\pm$0.01\\
\hline
\end{tabular}
\end{center}
\caption{"Ensemble" fractional variability (F$_{var}$) for NGC 925 ULX-1, ULX-2, ULX-3.  
The reported uncertainties derive from the standard deviation of the distribution of the mean F$_{var}$: see text for more details. 
}
\label{tab:fvarULX}
\end{table}

\begin{figure*}

 \includegraphics[width=19cm, height=7cm, angle=0]{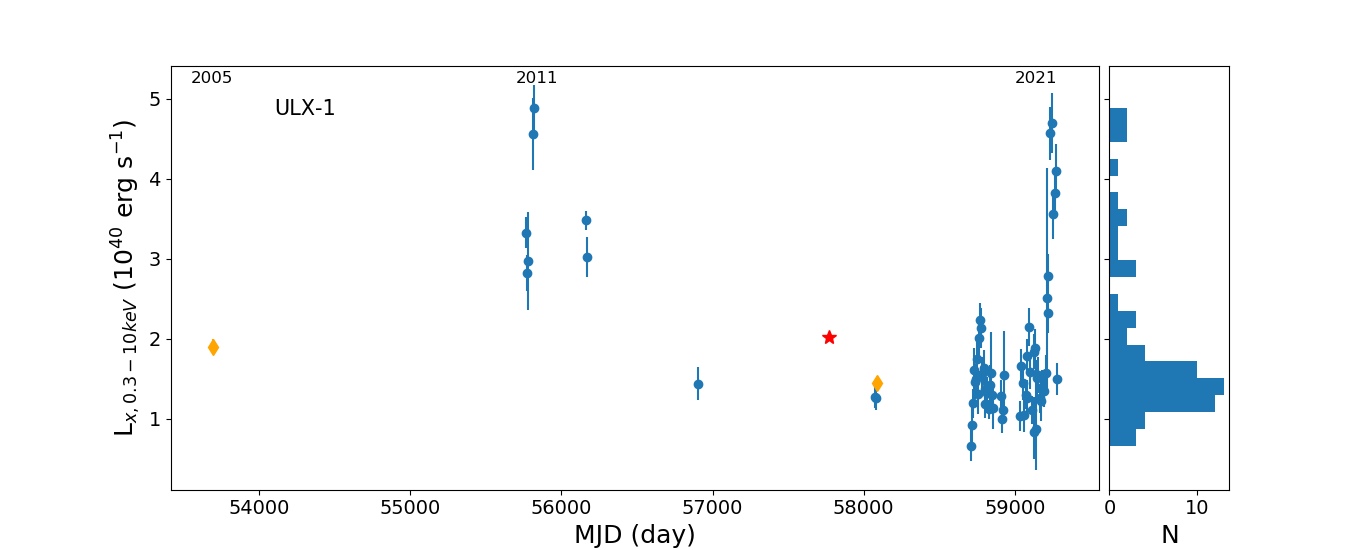}
\includegraphics[width=19cm, height=7cm, angle=0]{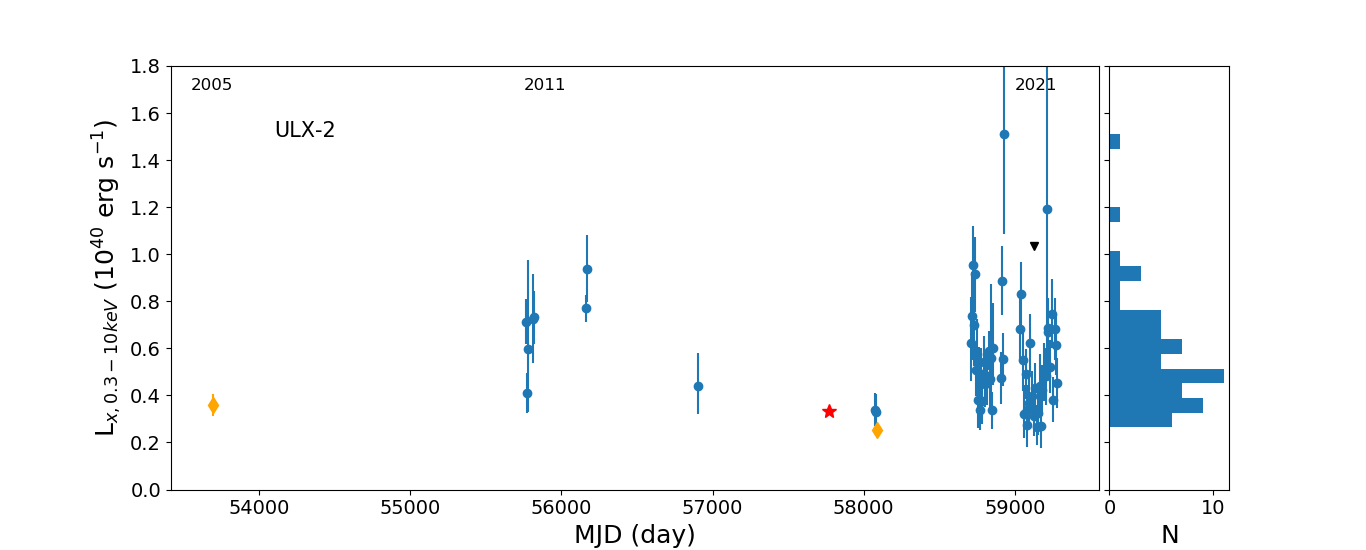}
\includegraphics[width=19cm, height=7cm, angle=0]{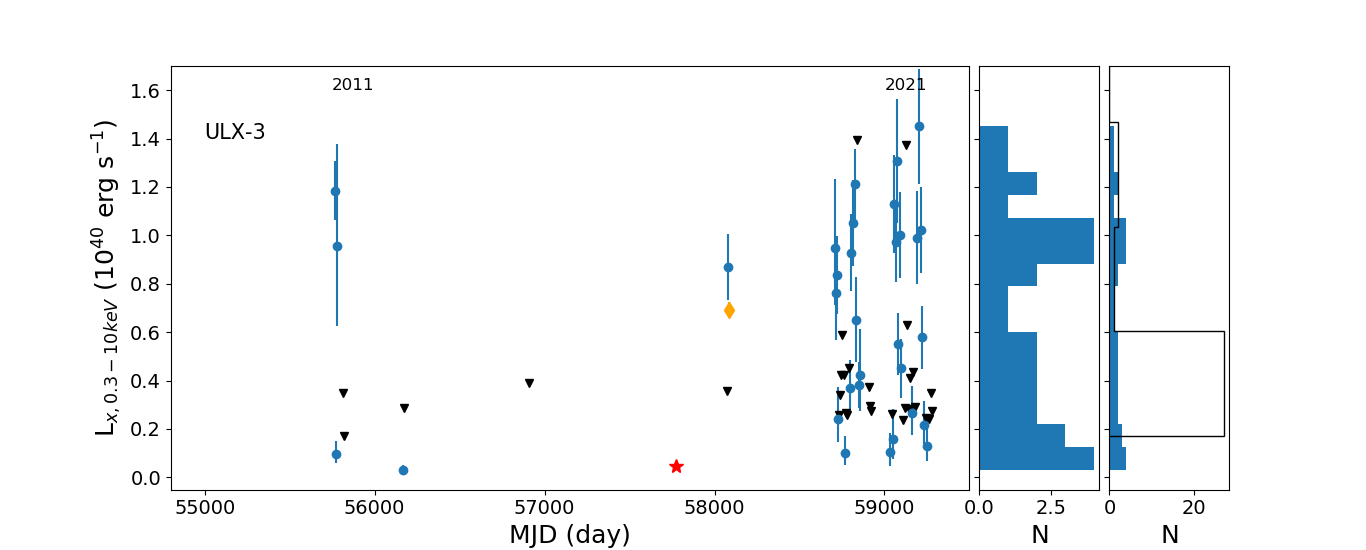}
 \caption{Light-curves, shown with six-day bins, and luminosity distributions of the three ULXs in NGC 925. The unabsorbed luminosity in the band [0.3-10] keV, in units of 10$^{40}$ erg s$^{-1}$, is derived from the best-fitting model in table \ref{tab:specULX}. The \swift\ detections are marked with blue dots while the \swift\ upper limits with the black triangles; the \xmm\ observation is marked  with the red star; the \chandra\ data with the orange diamonds. Top: ULX-1. - Left, light-curve; right, histogram of the distribution of the \swift\ luminosity.
   Middle: ULX-2. - Left, light-curve; right, histogram of the distribution of the \swift\ luminosity.
   Bottom: ULX-3. - Left, light-curve (note that the X-axis scale is different from that used in the other panels); middle, histogram of the distribution of the \swift\ luminosity; right, histogram of the distribution of the \swift\ luminosity, with the upper limits superimposed in black.}
   \label{Fig:ULX-light-curves}
\end{figure*}

We constructed the long-term \swift\ light-curves in the total energy band (0.3-10.0 keV) for each ULX and, for completeness, we added also the \xmm\ and \chandra\ data (Fig.~\ref{Fig:ULX-light-curves}); the conversion factor between \swift\ count-rate and luminosity was obtained by the spectral fit of the average \swift\ spectrum with the black body plus multi-colour disc model. The \chandra\ and \xmm\ luminosities have been derived from the spectral fit with the {\sc bbodyrad + diskbb} model. We also plot in a separate panel of Fig.~\ref{Fig:ULX-light-curves} the luminosity distribution for each ULXs, to look for any possible systematics, such as bi-modality.  
The \swift\ light-curves of all the sources are variable in flux on time-scales of at least days to weeks. We applied the Cash statistics \citep{Cash79} to verify at which significance the light-curves differ from a constant value. We also defined a flux variability factor for each ULX as the ratio of the highest to lowest flux. All these results are reported for each source in the next sections.

A classical method to quantify the flux variability, used e.g. to study sparsely sampled light-curves of accreting objects, such as AGNs (e.g. \citealt{Vagnetti2016,Aleksic2015,Schleicher2019}), is the fractional variability amplitude (F$_{var}$; e.g. \citealt{vaughan03,edelson2002}), which is defined as 
\begin{equation}
F_{var} = \frac{1}{\langle X\rangle}\sqrt{S^{2}-\langle\sigma_{err}^{2}\rangle}, 
\label{eq:Fvar}
\end{equation}
where S$^{2}$ is the total variance of the light-curve, $\langle \sigma_{err}^{2}\rangle$ is the mean error squared and $\langle X\rangle$ is the mean count rate. Here we use F$_{var}$ to study the long-term variability of ULXs. 

\citet{Allevato2013} demonstrated that the normalised excess variance evaluated from an individual light-curve, especially in case of irregularly sampled light-curves, might not be reliable. Thus, instead of deriving the estimate from the observed curve only, we used an ``ensemble'' approach, based on multiple simulated light-curves of the same source, as proposed in the cited paper. Since F$_{var}$ is simply the square root of the normalised excess variance, we applied the method proposed in \citet{Allevato2013} to our case. We performed a simulation of 5000 light-curves, starting from the properties of the observed one. We derived the F$_{var}$ applying the definition given in Equation \ref{eq:Fvar} for each simulated light-curve  
and grouped the obtained values in bins of 50 points. We evaluated a mean F$_{var}$ for each bin using Equation 12 in \citet{Allevato2013} and constructed the distribution of the obtained values. The estimates for F$_{var}$ and its uncertainty are the mean of the distribution and its standard deviation.  
We report the obtained values in Table \ref{tab:fvarULX}. We note that \citet{Allevato2013} also determine a bias on the variance estimate which depends both on the sampling pattern and on the power spectrum slope, if red noise is present in the data, as it may be in accreting compact objects (e.g. \citealt{vaughan03}). The red noise is characterised by a power spectrum with a power-law shape, with index 1--2 (e.g. \citealt{Press1978}). If red noise were present in our data with a slope $\sim$ 2, the estimate of the fractional variability may reduce of $\sim$ 25\%, while, for a slope $\sim$ 1, the effects coming from the irregular sampling pattern would dominate on the red noise and F$_{var}$ may increase of $\sim$ 10\%.
The derivation of F$_{var}$ does not take into account the upper limits. The effect of exclusion of significant upper limits is to underestimate F$_{var}$. Therefore, in such a case, the derived values have to be considered as lower limits on F$_{var}$. This happens mainly for ULX-3, where a number of significant upper limits is present in the light-curve, but not for ULX-1 and ULX-2, where the upper limits are rare.

We performed the simulations using a Monte Carlo approach. For each detection in the real light-curve, we extracted a random value from the normal distribution centered on the net (i.e. background-subtracted)  detected count rate and standard deviation as large as the error on the count rate. We repeated the same procedure for the observed background count rate of each observation, normalised for the ratio of source and background extraction area. We then summed the two count rates and multiplied them by the exposure time, in order to obtain the total counts in the source region. From the total counts we derived the statistical uncertainty on the simulated rate, using the Gehrels approximation \citep{Gehrels1986} when the total counts were less than 20, otherwise we used the square root of the counts. 

We also looked for possible periodicities in the long-term light-curve of each source. We applied the Lomb-Scargle approach (\citealt{Lomb1976,Scargle1982}, see also \citealt{vanderplas2018} for a recent review of the method) to derive the periodograms. We used the class \textit{LombScargle} of the subpackage \textit{timeseries} of the Python package for astronomy \textit{astropy}\footnote{\url{https://docs.astropy.org/en/stable/timeseries/lombscargle.html}}. 
We found a periodicity only for ULX-3. To better constrain the result we followed the same Monte Carlo approach used for the determination of the fractional variability. In this case, we considered also the upper limits. For each upper limit in the real light-curve we generated a random number from a uniform distribution, from 0 to the upper limit value and the background rate as done in case of source detection. We applied the same procedure used for the detections to derive the uncertainty.

\section{Results}
\label{sec:results}
\subsection{ULX-1}
\label{sec:results_ULX-1}
\begin{figure}
\includegraphics[width=6cm,angle=-90]{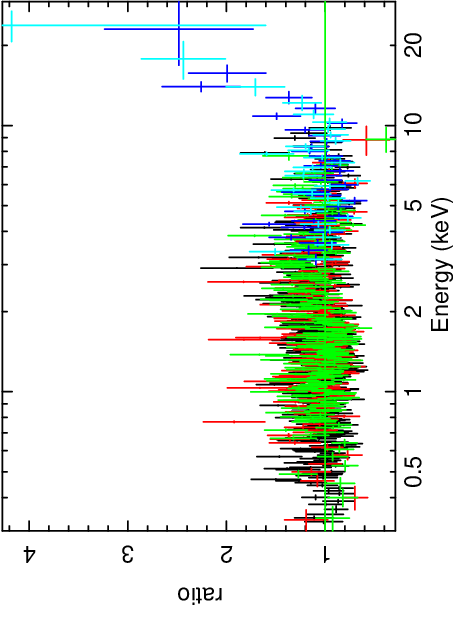}
\includegraphics[width=6cm,angle=-90]{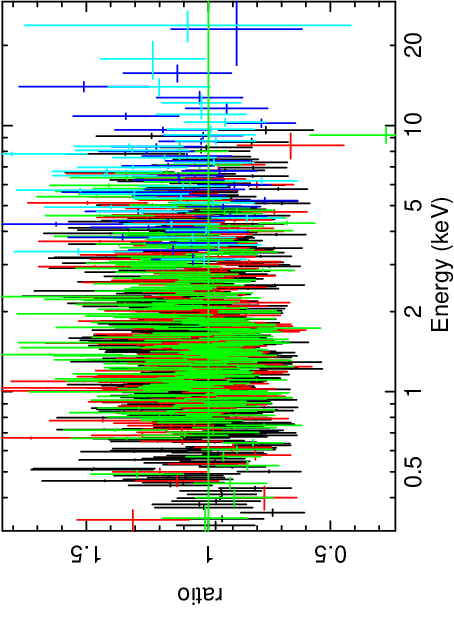}

   \caption{Upper: ULX-1 \xmm\ EPIC pn(black), MOS1 (red), MOS2 (green), \nus\ FPMA (blue), FPMB (cyan) residuals (data/model), for the {\sc bbodyrad} + {\sc diskbb}: an excess is clearly visible above 10 keV. Lower: ULX-1 residuals with the model {\sc bbodyrad} + {\sc diskbb} + {\sc cutoffpl}, the colours are the same used in the upper panel.} 
   \label{Fig:residui_mos_ulx2}
\end{figure}

ULX-1 is the most luminous of the three ULXs in NGC 925, with  peak (unabsorbed) luminosity of $\sim$ 5$\times$10$^{40}$ erg s$^{-1}$ in the \swift\ data
in 0.3-10 keV band. Its luminosity is in the range of the $\sim$ 10\% most luminous ULXs presently known (e.g. \citealt{swartz2011,Waltn2011,Earnshaw2019,Walton2021,Bernadich2021}). The significance of its temporal variability is 
$\sim$ 37$\sigma$, calculated using the Cash statistics \citep{Cash79} on the timescale of $\sim$ 6 d -- 10 yr. ULX-1 has a F$_{var}$ of $\sim$ 54\% in the total energy band  
and we estimated a flux variability as high as a factor of $\sim$ 8.
The fit of the \xmm\ / \nus\ spectrum, modelled with the {\sc bbodyrad} + {\sc diskbb} model, leaves some residuals above 10 keV (see Figure \ref{Fig:residui_mos_ulx2} upper panel), suggesting the need for a third spectral component. A possibility is to add a cutoff power-law component ({\sc cutoffpl} in {\sc xspec}) as found in other ULXs (e.g. \citealt{Walton2018b}). We obtained kT$_{bbodyrad}$ = 0.27$^{+0.02}_{-0.02}$ keV, kT$_{diskbb}$ = 1.3$^{+0.6}_{-0.2}$ keV, $\Gamma$ = 0.1$^{+0.6}_{-2.7}$ and HighECut = 5.2$^{+1.9}_{-3.4}$ keV, with $\chi^{2}$/dof = 1288.3/1267. 
We fixed the n$_{\rm H}$ to the best-fitting value obtained by modelling \xmm\ data only, because the more complex model did not allow us to constrain it. The residuals of the fit are shown in the bottom panel of Fig. \ref{Fig:residui_mos_ulx2}. This additional spectral component describes well the spectrum above 10 keV, but other models are possible given the low statistics in the \nus\ data. 
The \swift\ spectrum of the source is well described by a black body component with temperature kT$_{bbodyrad}\sim$ 0.3 keV and a multi-colour disc with inner temperature of kT$_{diskbb} \sim$ 2.1 keV. 

The spectral parameters in the (0.3--10 keV) band found for the observations taken with the different satellites are consistent, despite the large flux variability, suggesting the absence of marked spectral variability. 

From the HID (Figure \ref{Fig:hardness_hard_su_tot}-top), we note that there is no overall correlation. However, if we consider only the rates lower than 0.04 cts s$^{-1}$ (i.e. the three average bins at lower count rates), the linear fit gives a large correlation coefficient of 0.98. The regression model is plotted in Figure \ref{Fig:hardness_hard_su_tot} (upper panel).

\subsection{ULX-2}
\label{sec:results_ULX-2}
ULX-2 is the least variable in flux of the three ULXs and its peak luminosity (unabsorbed) in the \swift\ data is $\sim$ 1.5$\times$10$^{40}$ erg s$^{-1}$. The significance of its variability is $\sim$
8$\sigma$ and the flux variability factor is $\sim$ 5. The fractional variability is dominated by the hard band (F$_{var} \sim$59\%). Instead, in the soft band the F$_{var}$ results to be only $\sim$26\% (see Table \ref{tab:fvarULX}). 

From the spectral analysis of the \swift\ stacked spectrum, we found that the {\sc bbodyrad} temperature is $\sim$ 0.03 keV, but its normalization is not constrained, indicating that such a component is not significantly requested by the data. In fact, fitting the \swift\ average spectrum with a single {\sc diskbb}, we found a fit statistics similar to that obtained with the two components model. The {\sc diskbb} component temperature converges at $\sim$ 1.3 keV. Also the \chandra\ spectra can be modelled with a single component, while for the \xmm\ data the {\sc diskbb} does not provide a statistically acceptable fit, suggesting that the need for the two components arises only when the statistics is high. The spectral properties observed in  the \xmm\ / \nus\ and \chandra\ spectral-fitting are consistent within 90\% uncertainties, while those inferred in the \swift\ spectrum are different (kT$_{diskbb}^{\xmm\ / \nus\ }$ = 1.9 keV vs kT$_{diskbb}^{\swift\ }$ = 1.3 keV and (kT$_{bbodyrad}^{\xmm\ / \nus\ }$ = 0.27 keV vs kT$_{bbodyrad}^{\swift\ }$ = 0.03 keV), indicating a possible spectral variability for ULX-2.

We found that the \swift\ count rate of ULX-2 is always smaller than 0.04 cts s$^{-1}$ in the 0.3--10 keV band, and the hardness ratio shows evolution with the rate. In particular, we determined a linear trend by using the binned data (correlation factor $\sim$ 0.93), which indicates that the source becomes harder at the highest rates.

\subsection{ULX-3}\label{sec:analysisULX-3}
ULX-3 is a transient ULX with hints of (super-)orbital periodicity \citep{Earnshaw2020}. According to the \swift\ long-term monitoring, its unabsorbed luminosity reaches a peak of $\sim$ 1.5$\times$10$^{40}$ erg s$^{-1}$, its variability significance is 
$\sim$22$\sigma$ and its flux variability factor is $\sim$ 7. However, the latter is only a lower limit because of the presence of many upper limits in its long-term light-curve: indeed, considering also the lowest Chandra upper-limit, \citet{Earnshaw2020} found a factor of variability up to $\sim$ 30. 

The $F_{var}$ of ULX-3 is $\sim$ 63\% in the total 0.3--10 keV energy band. The highest amount of variability is seen in the soft band ($<1.5$ keV) unlike what is usually found in ULXs, for which the hard band is generally the most variable, at least on short timescales \citep[e.g.][]{Sutton2013,Pintore2020}. However, such a behaviour is possibly due to the fact that here the spectral shape changes around 1 keV. Therefore, we repeated the estimate of the variability dividing the bands at 1 keV. In this case we found a larger value for F$_{var}$ in the hard energy band ($\sim$54\%), while in the soft band the variability is $\sim$40\%. This implies that the highest variability concentrates around 1-2 keV.

We analysed the average \swift\ spectrum of ULX-3 with the same model used for the other ULXs (i.e. {\sc bbodyrad + diskbb} in {\sc xspec}), obtaining a cool black body (kT$_{bbodyrad} \sim$ 0.1 keV) and a {\sc diskbb} with a temperature of $\sim$1.5 keV.

The hardness ratio of ULX-3 is consistent with a constant within the uncertainties, with a mean value of $\sim$ 0.5 at the detected count rates.

We applied the Lomb-Scargle method to the ULX-3 light-curve to search for periodic variability. We used  only the \swift\ observations, stacked in time bins of $\sim$ 6 d, taken between 2019 and March 2021 (i.e. MJD > 58500), where the monitoring was denser and more regular. This allowed us to exclude periods of long-term gaps where the behaviour of the source was unknown (for example it could have faded).

\begin{figure}
\begin{center}

\includegraphics[width=9.2cm, angle=0]{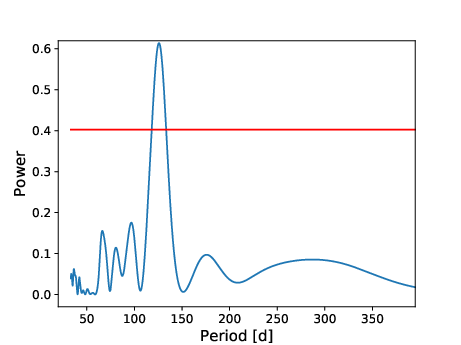}
\caption{(Super-)orbital period in NGC 925 ULX-3. Power vs. frequency, from the Lomb-Scargle periodogram. The level corresponding to 3$\sigma$ significance is indicated by the red line. }
      \label{Fig:ULX-timing}
\end{center}

\end{figure}

We applied the Lomb-Scargle approach to the frequency range [2$\times$10$^{-3}$ -- 0.08] d$^{-1}$, which corresponds to the period range 12--500 d. At first, we used only the \swift\ data with a source detection. We found several peaks in the power density spectrum, the most prominent of which corresponds to a period of $\sim$ 126 d. However, the peak was not statistically significant 
because of the sparse distribution of the data. Thus we decided to include also the information carried by the significant upper limits (UL), to reduce the gaps between the observations. We included them somewhat arbitrarily at half the value of each UL with an associated uncertainty of the same amount. 
The new Lomb-Scargle power spectra is showed in Fig.~\ref{Fig:ULX-timing}. The peak at the period of 126 d is now predominant.

\begin{figure*}
\begin{center}

\includegraphics[width=18cm, angle=0]{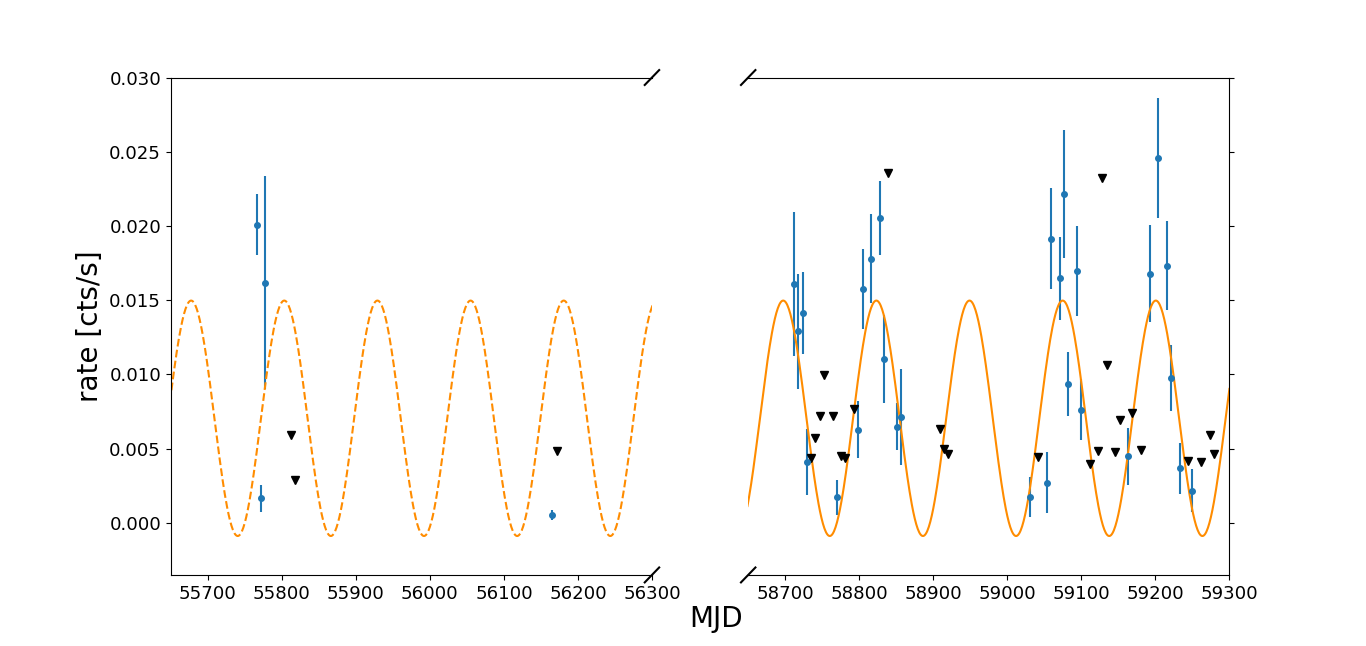}

   \caption{(Super-)orbital period in NGC 925 ULX-3. Light-curve binned in interval of 6 days of NGC ULX-3 with the sinusoidal best-fitting of the periodicity superimposed. The period has been estimated on the more recent and denser data (solid orange line) and has been extrapolated to the older data (dashed orange line). The blue dots are the \swift\ detections, while the black inverted triangles are upper limits.}
      \label{Fig:lc_period_ULX3}
\end{center}
\end{figure*}

To assess the goodness of our approach, we performed a Monte-Carlo simulation (see section \ref{sec:timing-analysis} for details) of 5000 light-curves considering both detections and upper limits, between 2019 and March 2021, 
in the original light-curve showed in Fig.~\ref{Fig:ULX-light-curves}-lower panel. In the simulations, the value corresponding to each upper limit is derived from a random value in the uniform distribution between zero and the upper limit value, considering each value equally probable. 
For each simulated light-curve 
we applied a Lomb--Scargle analysis. We constructed the histogram of the highest peak frequencies obtained from the simulated light-curves and we fit such a distribution with a Gaussian function. We found that the mean value and the standard deviation of the Gaussian are 126.1 d and 2.0 d, respectively. The best-period is therefore totally compatible with the one we found using only the \swift\ detections. Therefore we claim the existence of a periodicity of $126 \pm 2$ d. 
A superposition of the expected periodicity on the observed data is shown in Fig.~\ref{Fig:lc_period_ULX3}. 
If we extrapolate the detected periodicity to the data taken before 2019, we find that the curve does not follow the observations. This could indicate that either the periodicity has slightly changed or its phase has shifted in time, or it could depend on the uncertainties on the highest peak (see figure \ref{Fig:lc_period_ULX3}). A better superposition of the same periodicity on the older points is found with a shift in phase of $\sim$ 55 days in about 10 years,
although the number of points is small: similar shifts have been observed in the super-orbital periodicity of other ULXs, see for example \citet{Brightman2019}.

The significance of the peak corresponding to the best period has been determined with the {\it Baluev} method \citep{Baluev2008}, an analytical approximation to determine the false alarm probability (FAP), i.e. the probability to obtain a peak equal or higher than the considered one if only noise is present. From the observed light-curve, we obtained a FAP value of $\sim$ 10$^{-8}$, which should be considered as an upper limit. Therefore the significance of the peak should be $>$5$\sigma$.
We must notice that the determination of the peak significance through the FAP assumes white noise (the height of spurious peaks is independent from frequency; \citealt{vanderplas2018}). This assumption could be not true for compact accreting objects, which may present red noise \citep{vaughan03}. To take into account this effect, we used an approach similar to that followed by \citet{Walton2016} for NGC 5907 ULX-1. We simulated 5000 light-curves with a red noise power spectrum, assuming a power-law with slope 1 and 2. We performed the simulations using the \textit{Simulator} object of the Python library for spectral timing \textit{Stingray} v.0.2\footnote{\url{https://docs.stingray.science}; \citealt{matteobachetti2020}}, which applies the algorithm from \citet{Timmer1995} (see \citealt{Huppenkothen2019} and \citealt{Huppenkothen2019_openjournal} for a description of the library). We simulated continuous light-curves, with a time resolution of 2 ks, and we selected only time bins corresponding to the epochs of the real data. For each simulated curve, we ran the Lomb--Scargle method to look for peaks comparable to the real data one but generated by random noise. 
In the worst case, i.e. slope=2, we found that most of the peaks generated from the random noise have frequencies smaller than 0.0035 1/d (280 d), which corresponds to roughly half of our observing window. Therefore peaks corresponding to longer periods (i.e. smaller frequencies) are about the 70\% of the total number of peaks. However, these peaks are not reliable because they would cover less than 2 cycles in the light-curve time window. Only two-three peaks have a period comparable or smaller and a significance larger than that of the real peak from the observed light-curve. These correspond to $\sim$ 0.05\% of the total number of peaks derived from the red noise simulations. Therefore the significance of the periodicity detection may decrease to $\sim$ 3.4$\sigma$ in the worst case of possible red noise.  For a slope=1, only two peaks in the total frequency band have significance larger than our period detection, both of them at significantly smaller frequencies than that of the periodicity. So, even if red noise were present, it is unlikely that the peak corresponding to the period in the real data has been generated by chance from the random noise.

We also wanted to verify if a short-term pulsation period, such as those found in the PULX, was present. So we searched for coherent signals in ULX-3 \xmm\ data also including the correction for a possible first period derivative component (1$\times$10$^{-6} < \left| P_{dot}/P \right| < 1\times$10$^{-12}$) but no significant signals were detected. The 3$\sigma$ upper limits on the fractional amplitude (defined as the semi-amplitude of the sinusoid divided by the average source count rate) are between 70\% $-$ 100\% in the period range 500$-$0.150s

\section{Discussion}
\label{sec:discussion}

We have studied the timing and spectral properties of the three ULXs in the galaxy NGC 925, analysing all the available \swift, \xmm, \chandra\ and \nus\ public observations. In the following, we discuss the main results of this manuscript and their implications for the detected ULXs.

\subsection{Temporal variability}
The three ULXs in NGC 925 are confirmed to be all variable in flux on days-to-weeks timescales. The fractional variability has been often used to study the variability properties of ULXs within individual observations, typically considering timescales of minutes -- hours (e.g. \citealt[]{Sutton2013,Pintore2020,Mondal2021,Robba2021}). Here, we used this estimator on longer timescales (using 6 days bins), applying it to sparse sampled light-curves, using the `ensemble' approach proposed in \citet{Allevato2013}, as explained in section \ref{sec:timing-analysis}.

The most luminous of the ULXs in NGC 925 is ULX-1, which alternated epochs of "high-flux" (0.3-10 keV  $F_{abs} > 3\times 10 ^{-12}$ erg s$^{-1}$ cm$^{-2}$ ) in some of the observations taken in 2011, and "low-flux" (0.3-10 keV  $F_{abs} < 3\times$10$^{-12}$ erg s$^{-1}$ cm$^{-2}$) in the subsequent observations. However, from February 2021 the source entered again in a high flux regime, 
from which it dropped on 8 March 2021, the last observation before a period of non-visibility for the \textit{Swift} satellite that was monitoring the galaxy. It has a large F$_{var}$, with a value of $\sim$54\% in the total energy band. Similarly, ULX-3 was also highly variable (F$_{var}\sim60\%$) on weekly timescales.

ULX-2 is the least variable of the three ULXs, with a variability amplitude of $\sim$ 35\% in the total band. 
The variability is driven by the high energy component, with $F_{var}$ $> 50\%$, while the soft band variability is just 20-30\%. This trend is also present in the other two ULXs, even if it is less pronounced. 
This behaviour is consistent with the results of previous works which analysed other ULXs variability on smaller time-scales \citep[e.g.][]{middleton2015b,Walton2018a}. 

The transient ULX-3 is the only one of the three ULXs in NGC 925 to show a candidate (super-)orbital periodicity.  
From the Lomb-Scargle method,  we estimated a periodicity consistent with the interval of 70-150 days previously found by \citet{Earnshaw2020}. In particular, we found that, during the \swift\ monitoring campaign which started in 2019, the source presents a periodicity of $\sim126$ d, which we proved to be significant. 
The observations used cover only four cycles of the inferred periodicity, and we found that an extrapolation of it at earlier epochs is not consistent with the data. Either the source experienced a period of turn-off of the accretion during the epochs where it was not monitored, or the phase of the periodicity may have changed for other effects as the onset of a propeller regime or a change in the geometrical configuration of the accretion system. Further coverage of the source should give better constraints.

Periodicities of the order of tens/hundreds of days discovered from the flux variability in ULX light-curves are often interpreted as super-orbital variabilities (e.g. \citealt{Israel2017}, \citealt{Brightman2019}) and may be linked to some kind of precession, e.g. of a warped accretion disc, of a large height disc or a Lense-Thirring precession, as has been found in many binaries (e.g. \citealt[]{motch2014,Hu2017,Fuerst2017,Middleton2018}).

Long-term light-curves of ULXs can be used to search for possible propeller phases (e.g. \citealt{Earnshaw2018, Song2020}), indicating the presence of a NS accretor. Given that, during a propeller phase, the accretion on the compact object is expected to stop, the presence of a flux bi-modal distribution with orders of magnitude difference, corresponding to active and non-active periods of accretion, identifies PULX candidates.
Despite its large flux variability, there are no indications of propeller phases in the light-curve of ULX-1: the histogram of the luminosity (figure \ref{Fig:ULX-light-curves}, upper panel - right) does not indicate a bi-modal distribution.  
Furthermore, \citealt{Lara-Lopez2021} found a low-metallicity environment around ULX-1, which in principle could favour the formation of BHs with respect to NSs, since less mass loss is expected in low-metallicity stars (e.g. \citealt{Heger2003}). In contrast, the excess found in the residuals above 10 keV in the \nus\ data may be explained with the emission of the accreting column above the magnetic poles of a NS, as seen in other ULXs (e.g. \citealt{Walton2018b}). 
Also for ULX-2 the luminosity is always larger than 10$^{39}$ erg s$^{-1}$ and no bi-modal flux distribution is observed in the light-curve. ULX-3 is the only ULX in NGC 925 with many upper limits. Between MJD 56165 and 57770 (about 4 years) the source has been observed only twice with \textit{Swift}/XRT, without detections, before the one with \xmm\ (MJD 57771). We cannot exclude that in this period the source flux could have decreased by a large amount, because of a propeller phase, but the upper limits found are also compatible with the lower fluxes of the (super-)orbital modulation. \citet{Lara-Lopez2021} found high-metallicity around ULX-3, consistent with the possibility that the compact object may be a NS. The confirmation of the presence of a NS would come from the detection of pulsations, but the low statistics of the \xmm\ data allowed us to just derive an upper limit on the fractional amplitude (see section \ref{sec:analysisULX-3}). Only upper limits have been found also for the pulsation period in ULX-1 and ULX-2 (see \citealt{Pintore2018b} for details).

\subsection{Spectral classification}
Following the classification described in \citet{Sutton2013} and using the 2017 \xmm\ data, we classified the spectral regime of the three ULXs (see also section \ref{sec:introduction} for details on the classification method). 

ULX-1 was spectrally stable in all the available data and, according to the best-fitting {\sc diskbb+powerlaw} model ($\Gamma \sim$ 1.8, T$_{diskbb} \sim$ 0.2), ULX-1 can be classified as a \textit{hard ultraluminous} source (HUL). However, considering also the \nus\ data the model leaves an excess at high energies, suggesting a spectral curvature (as found by \citealt{Pintore2018b}), which resembles the spectral shape of a \textit{broadened disc} (BD). \citet{Sutton2013} highlighted the difficulty to distinguish the HUL regime from the BD one, for sources with a pronounced curvature, perhaps caused by a strong comptonization. In order to model the high energy spectral curvature we fit the \xmm+\nus\ spectrum of ULX-1 with a multi-colour disc (MCD) plus a high energy cutoff power-law ({\sc highecut$\times$powerlaw}, in {\sc xspec}). The $\Gamma$ and kT values we found are consistent with the ones from the previous fit with \xmm\ data only, confirming the classification as an HUL source for ULX-1. Similar results were also confirmed by the \swift\ average spectrum. This indicates that, despite the flux variability, there is no evident spectral variability over the considered period. The HID of ULX-1 confirms that it is a rather hard source. \citet{Pintore2018b} proposed that the hard emission from ULX-1 could be explained by a system seen at a small inclination angle (nearly face-on), where, in a scenario of super-Eddington accretion onto a stellar mass compact object, the hard emission from the innermost regions of the accretion flow is not covered by the outflows. They also studied the hardness ratio for NGC 925 ULX-1, using a colour-colour diagram, in the energy band 2-30 keV and found that the position of ULX-1 was close to that of the PULXs analysed in \citet{Pintore2017}. With the available data analysed here, we could not determine the nature of the compact object in ULX-1, which remains still unknown.

The fit of the simultaneous \xmm+\nus\ spectra of ULX-2 with a multi-colour disc plus a power-law (kT=0.2 keV and $\Gamma$=1.9) allowed us to classify ULX-2 as likely a hard Ultraluminous (HUL) source. 
However, if we take into account the uncertainties on $\Gamma$, which can vary between 1.7 and 2.1,  
the source could be a SUL ULX as well. As pointed out by \citet{Sutton2013}, both regimes can take place in the same system, depending on changes in the accretion rate and in the wind's opening angle, which has the shape of a funnel (as resulted from the simulations of \citealt{Kawashima2012}). So we suggest that ULX-2 is seen at intermediate inclination angles (between face-on direction, where the hard emission from the inner regions would be predominant, and the funnel inclination angle direction, where we would see the softer emission of the wind's photosphere) appearing in a SUL/HUL regime. We found a linear correlation, with the source becoming harder at larger count rates, in the HID of ULX-2. 
This could be interpreted with a source seen at small inclination angles, nearer to the face-on position than to the edge-on one, leaving the inner regions of the accretion flow, which become brighter and hotter with increasing rates, directly visible to the observer. An alternative explanation is linked to the beaming. While the count rate increases, due to an enhancement in the mass accretion rate ($\dot{m}$), the wind opening angle narrows. The hard emission becomes consequently more beamed. Therefore both the hard and soft emission increase, but the beamed hard emission increase is proportional to $\dot{m}^{2}$ (\citealt{King2009}), while the soft one to $\dot{m}$, resulting in an overall hardening (see e.g. \citealt{middleton2015a}). This mechanism has been proposed by \citet{Luangtip2016} to explain the source Ho IX X-1, where it was also observed a shift of the peak in the hard component towards lower temperatures, while the luminosity increases. The authors explained this considering the simulated spectra for comptonising BHs (\citealt{Kawashima2012}), which appear to be harder at lower super-Eddington rates, because of photon up-scattering in the shocked region near the BH and softer at extreme super-critical rates, when the wind funnel narrows reducing the number of photons which can escape without intercepting the wind and being Compton down-scattered.

For ULX-3 we found a cool disc temperature ($\sim$ 0.09 keV) and a photon index between $\sim$ 1.7 and 3.2, so in this ULX may be present both the HUL and SUL regimes.
The average \swift\ spectrum (magenta in the lower panel of figure \ref{Fig:unfold_spec}) of ULX-3, which is more luminous than the \xmm\ one, has also a harder spectral shape, with best-fitting values kT$_{diskbb} \sim$ 0.09 keV and $\Gamma \sim$ 1.6. From this data ULX-3 can be classified as a \textit{hard ultraluminous} ULX. This trend is in accordance with the result of \citet{Earnshaw2020}, who found a softer spectrum at low luminosities which became harder at higher ones. As already noted by \citet{Earnshaw2020}, this behaviour could indicate a propeller effect causing a soft and dim state in the absence of accretion onto the compact object, but the flux observed in the \xmm\ observation is still compatible with the periodic flux modulation. In contrast, the hardness ratio seems constant among the observed count rates, within the errors. 

Considering the known PULXs, they are usually found in the \textit{hard ultraluminous} regime (e.g. \citealt{Sutton2013,Pintore2017,Carpano2018}). The disc temperatures of the ULXs in NGC 925 are similar to those observed in the PULXs ($\sim$ 0.04 -- 0.2 keV),  
while the power-law index of the PULXs ($\Gamma\sim1-1.9$) are sometimes comparable to and other times harder than the sources analysed in this work.

\subsection{Super-Eddington accretion}

From the normalizations of the fitting components, it is possible to derive the apparent emitting radii for the {\sc bbodyrad} and {\sc diskbb} components. We remark that such dimensions might not have a real physical meaning but they allow us to give a first, tentative characterisation of the accretion geometry. 

From the \swift\ average spectrum of ULX-1, the radii of the {\sc bbodyrad} and {\sc diskbb} are 1452$^{+588}_{-362}$ km and 75$^{+14}_{-17}$ km respectively (assuming a reference inclination angle of 45$^{\circ}$). Such values are similar to those observed in a number of ULXs (e.g. 4X J1118, \citealt{Motta2020}; Circinus ULX-5, \citealt{Mondal2021}; \citealt{Walton2020}). Considering that the source is in an ultraluminous accretion regime, we expect the accretion disc to be supercritical, with advection (slim disc model) and outflows (e.g. \citealt{Shakura1973,Poutanen2007}). The disc is expected to be geometrically thick inside the spherisation radius\footnote{The radius at which the vertical component of gravity becomes comparable with the force exerted by the radiation pressure},   
where a fraction of the accreting matter is radiatively expelled from the disc in the form of a powerful wind, and geometrically thin in the very outer regions. The latter can be cold and not contribute much to the total ULX flux. In addition, the inner regions of the accretion flow may be also naked, i.e. the outflows are no more present above the thick disc. As proposed by some authors (e.g. \citealt{Walton2015b,Pintore2018b}),  
in this scenario the larger radius could be interpreted as an average size of the region where the winds are launched, while the smaller radius may be the size of the inner disc. Assuming accretion onto a non-rotating BH, the inner disc radius (R$_{in}$) would be three times the Schwarzschild radius, i.e. R$_{in}$ = 6 GM / c$^{2}$. The apparent inner disc dimension of $\sim75$ km is too small to be the inner disc radius of an IMBH (which would result larger than $\sim900$ km, assuming a minimum mass of 100M$_{\odot}$), but also of a massive stellar mass BH (30--100 M$_{\odot}$), suggesting that the ULX may host a lighter stellar mass BH (M$_{BH} <$ 10M$_{\odot}$) or a NS. 

The radius of the hot component derived from the \swift\ average spectrum for ULX-2 is 89$^{+12}_{-14}$ km (for an inclination angle of 45$^{\circ}$). This value is consistent with that found by \citet{Pintore2018b} from the \xmm\ / \nus\ data and the considerations for ULX-1 hold for ULX-2 also, i.e. the internal disc radius may correspond to that of a light BH or a NS.

In the case of ULX-3, the radius associated with the colder emitting region is 7200$^{+6100}_{-3000}$ km, while the hotter emitting region has a radius R$_{diskbb}$ = 47$^{+11}_{-14}$ km for a 45$^{\circ}$ inclination angle. The last one, if corresponding to the inner disc radius of an accreting compact object, would belong to a compact object mass even smaller than in the case of ULX-1 and ULX-2.

\section{Conclusion}
\label{sec:conclusion}
All the three ULXs analysed in this work are variable on time-scales of days/months. One of them, ULX-3, has a candidate (super-)orbital periodicity on a time-scale of months ($\sim$ 126 d) similar to the super-orbital periodicities observed in some other ULXs.

The spectral study of all the three sources suggests that they are all in a super-Eddington accretion regime, the \textit{hard ultraluminous regime} for ULX-1 and an intermediate situation between the \textit{hard} and \textit{soft ultraluminous regime} for ULX-2 and ULX-3, for which a unique classification is not possible, 
indicating that repeated observations are necessary to investigate fully the emission of ULXs. While ULX-1 shows only weak spectral variability, for the other two sources we found hints of a significant spectral variability. The spectral-timing properties suggest that ULX-1 and ULX-2 are being seen close to face-on, while ULX-3 may be seen through the wind in a super-Eddington scenario.

Follow-up observations would be useful to double check the periodicities and place stronger constraints on the pulsed fraction.

\section{Acknowledgements}
This work has been partially supported by the ASI-INAF program I/004/11/4 and the ASI-INAF agreement n.2017-14-H.0. EA acknowledges funding from the Italian Space Agency, contract ASI/INAF n. I/004/11/4. We acknowledge the usage of the HyperLeda database (http://leda.univ-lyon1.fr). This work made use of the data from the UK Swift Science Data Centre (University of Leicester) and of the XRT Data Analysis Software (XRTDAS) developed by the ASI Science Data Center (Italy). 

\section{Data availability}
The Swift, XMM-Newton, Chandra and NuSTAR data used in this work are available respectively from {https://www.swift.ac.uk/swift\_portal/}, https://www.cosmos.esa.int/web/xmm-newton/xsa, https://cxc.harvard.edu/cda/ and https://heasarc.gsfc.nasa.gov/.

\bibliographystyle{mnras}
\bibliography{main}{}

\bsp	
\label{lastpage}
\end{document}